\documentclass[trackchanges,twocolumn]{aastex701}
\usepackage{graphicx}
\usepackage{wrapfig}
\usepackage{caption} 
\usepackage{float}

\newcommand{\teff}{$T_{\rm eff}$}
\newcommand{\logg}{$\log g$}

\newcommand{\csi}{$\xi$}
\newcommand{\feh}{[Fe/H]}

\newcommand{\vsini}{$v\sin i$}

\begin{document}

\title{The Hot Neptune Initiative (HONEI):\\ IV. TOI-4129\,b, an Adolescent Neptune in the Ridge}
\shorttitle{A ``Nepteenian" planet in the Ridge}
 
\author[orcid=0009-0009-2072-0383]{Sureena Lakhanpal}
\affiliation{Department of Astronomy, University of California, Berkeley, CA 94720, USA}
\affiliation{The Observatories of the Carnegie Institution for Science, 813 Santa Barbara Street, Pasadena, CA 91101, USA}
\email[show]{sureena\_lakhanpal@berkeley.edu}
\author[orcid=0000-0003-2527-1475]{Shreyas Vissapragada}
\affiliation{The Observatories of the Carnegie Institution for Science, 813 Santa Barbara Street, Pasadena, CA 91101, USA}
\email{svissapragada@carnegiescience.edu}
\author[orcid=0000-0001-9390-0988]{Luca Naponiello}
\affiliation{INAF -- Osservatorio Astrofisico di Torino, Via Osservatorio 20, 10025 Pino Torinese (TO), Italy}
\email{luca.naponiello@inaf.it}
\author[0009-0000-5623-5237]{Simone Filomeno}
\affiliation{INAF -- Osservatorio Astronomico di Roma, Via Frascati 33, 00078 Monte Porzio Catone (RM), Italy}
\email{simone.filomeno@inaf.it}
\author[orcid=0000-0002-1892-2180]{Katia Biazzo}
\affiliation{INAF -- Osservatorio Astronomico di Roma, Via Frascati 33, 00078 Monte Porzio Catone (RM), Italy}
\email{katia.biazzo@inaf.it}
\author[orcid=0000-0002-9148-034X]{Vincent Bourrier}
\affiliation{Observatoire astronomique de l'Universit\'e de Gen\`eve, Chemin Pegasi 51, 1290 Versoix, Switzerland}
\email{vincent.bourrier@unige.ch}
\author[orcid=0000-0001-6880-5356]{Tiziano Zingales}
\affiliation{Dipartimento di Fisica e Astronomia ``Galileo Galilei'', Universit\`a di Padova, Vicolo dell'Osservatorio 3, 35122 Padova, Italy}
\affiliation{INAF -- Osservatorio Astronomico di Padova, Vicolo dell'Osservatorio 5, 35122 Padova, Italy}
\email{tiziano.zingales@unipd.it}
\author[orcid=0009-0002-2548-4948]{Erik Friden}
\affiliation{Observatoire astronomique de l'Universit\'e de Gen\`eve, Chemin Pegasi 51, 1290 Versoix, Switzerland}
\email{erik.friden@unige.ch}
\author[orcid=0000-0002-1533-9029]{Emily K. Pass}
\affiliation{Kavli Institute for Astrophysics and Space Research, Massachusetts Institute of Technology, Cambridge, MA 02139, USA}
\email{epass@mit.edu}
\author[orcid=0000-0002-6177-198X]{Aldo S. Bonomo}
\affiliation{INAF -- Osservatorio Astrofisico di Torino, Via Osservatorio 20, 10025 Pino Torinese (TO), Italy}
\email{aldo.bonomo@inaf.it}
\author[orcid=0000-0001-9911-7388]{David W. Latham}
\affiliation{Center for Astrophysics \textbar \ Harvard \& Smithsonian, 60 Garden Street, Cambridge, MA 02138, USA}
\email{dlatham@cfa.harvard.edu}
\author[orcid=0000-0001-6588-9574]{Karen A.\ Collins}
\affiliation{Center for Astrophysics \textbar \ Harvard \& Smithsonian, 60 Garden Street, Cambridge, MA 02138, USA}
\email{karen.collins@cfa.harvard.edu}
\author[orcid=0000-0002-5741-3047]{David R. Ciardi}
\affiliation{NASA Exoplanet Science Institute, Caltech/IPAC, 1200 E.\ California Blvd, Pasadena, CA 91125, USA}
\email{ciardi@ipac.caltech.edu}

\author[0000-0002-2970-0532]{David Baker} 
\affiliation{Physics and Engineering Department, Austin College, Sherman, TX 75090, USA}
\email{dbaker@austincollege.edu}
\author[0000-0003-1464-9276]{Khalid Barkaoui} 
\affiliation{Instituto de Astrof\'isica de Canarias (IAC), Calle V\'ia L\'actea s/n, E-38205, La Laguna, Tenerife, Spain}
\affiliation{Astrobiology Research Unit, Universit\'e de Li\`ege, 19C All\'ee du 6 Ao\^ut, 4000 Li\`ege, Belgium}
\affiliation{Department of Earth, Atmospheric and Planetary Science, Massachusetts Institute of Technology, 77 Massachusetts Avenue, Cambridge, MA 02139, USA}
\email{khalid.barkaoui@uliege.be}
\author[0000-0002-8863-7828]{Andrew Collier Cameron} 
\affiliation{SUPA School of Physics \& Astronomy, University of St Andrews, North Haugh, St Andrews KY19 9SS, UK}
\email{acc4@st-andrews.ac.uk}
\author[0000-0001-6319-9297]{Patrick Eggenberger} 
\affiliation{Observatoire astronomique de l'Universit\'e de Gen\`eve, Chemin Pegasi 51, 1290 Versoix, Switzerland}
\email{Patrick.Eggenberger@unige.ch}
\author[0000-0002-4909-5763]{Akihiko Fukui} 
\affiliation{Komaba Institute for Science, The University of Tokyo, 3-8-1 Komaba, Meguro, Tokyo 153-8902, Japan}
\affiliation{Instituto de Astrof\'isica de Canarias (IAC), Calle V\'ia L\'actea s/n, E-38205, La Laguna, Tenerife, Spain}
\email{afukui@g.ecc.u-tokyo.ac.jp}
\author[0000-0002-1715-6939]{Ancy Anna John} 
\affiliation{School of Physics \& Astronomy, University of Birmingham, Edgbaston, Birmingham, B15 2TT, UK}
\email{a.a.john@bham.ac.uk}
\author[0000-0002-5375-4725]{Heather A. Knutson} 
\affiliation{Division of Geological and Planetary Sciences, California Institute of Technology, Pasadena, CA 91125, USA}
\email{hknutso2@caltech.edu}
\author[orcid=0000-0003-3204-8183]{Mercedes L\'{o}pez-Morales} 
\affiliation{Space Telescope Science Institute, 3700 San Martin Drive, Baltimore, MD 21218, USA}
\email{mlopez-morales@stsci.edu}
\author[0000-0001-8342-7736]{Jack Lubin} 
\affiliation{Department of Physics \& Astronomy, University of California Los Angeles, Los Angeles, CA 90095, USA}
\email{jblubin@ucla.edu}
\author[orcid=0000-0002-9428-8732]{Luigi Mancini} 
\affiliation{Dipartimento di Fisica, Universit\`a di Roma ``Tor Vergata'', Via della Ricerca Scientifica 1, 00133 Roma, Italy}
\affiliation{INAF -- Osservatorio Astrofisico di Torino, Via Osservatorio 20, 10025 Pino Torinese (TO), Italy}
\email{lmancini@roma2.infn.it}
\author[orcid=0009-0003-1841-7381]{Francesca Manni} 
\affiliation{Dipartimento di Fisica, Universit\`a di Roma ``Tor Vergata'', Via della Ricerca Scientifica 1, 00133 Roma, Italy}
\affiliation{INAF -- Osservatorio Astrofisico di Torino, Via Osservatorio 20, 10025 Pino Torinese (TO), Italy}
\email{francesca.manni@inaf.it}
\author[orcid=0000-0002-6871-6131]{Giacomo Mantovan} 
\affiliation{Centro di Ateneo di Studi e Attivit\`a Spaziali ``G. Colombo'' -- Universit\`a di Padova, Via Venezia 15, IT-35131, Padova, Italy}
\affiliation{INAF -- Osservatorio Astronomico di Padova, Vicolo dell'Osservatorio 5, 35122 Padova, Italy}
\email{giacomo.mantovan@unipd.it}
\author[orcid=0000-0001-7254-4363]{Annelies~Mortier} 
\affiliation{School of Physics \& Astronomy, University of Birmingham, Edgbaston, Birmingham, B15 2TT, UK}
\email{a.mortier@bham.ac.uk}
\author[orcid=0000-0001-9087-1245]{Felipe Murgas} 
\affiliation{Instituto de Astrof\'isica de Canarias (IAC), Calle V\'ia L\'actea s/n, E-38205, La Laguna, Tenerife, Spain}
\affiliation{Departamento de Astrof\'isica, Universidad de La Laguna (ULL), E-38206 La Laguna, Tenerife, Spain}
\email{fmurgas@iac.es}
\author[0000-0001-8511-2981]{Norio Narita} 
\affiliation{Komaba Institute for Science, The University of Tokyo, 3-8-1 Komaba, Meguro, Tokyo 153-8902, Japan}
\affiliation{Astrobiology Center, 2-21-1 Osawa, Mitaka, Tokyo 181-8588, Japan}
\affiliation{Instituto de Astrof\'isica de Canarias (IAC), Calle V\'ia L\'actea s/n, E-38205, La Laguna, Tenerife, Spain}
\email{narita@g.ecc.u-tokyo.ac.jp}
\author[0000-0003-1360-4404]{Belinda~A.~Nicholson} 
\affiliation{Centre for Astrophysics, University of Southern Queensland, West St. Toowoomba, QLD Australia 4350}
\email{Belinda.Nicholson@unisq.edu.au}
\author[orcid=0000-0003-0987-1593]{Enric Palle} 
\affiliation{Instituto de Astrof\'isica de Canarias (IAC), Calle V\'ia L\'actea s/n, E-38205, La Laguna, Tenerife, Spain}
\affiliation{Departamento de Astrof\'isica, Universidad de La Laguna (ULL), E-38206 La Laguna, Tenerife, Spain}
\email{epalle@iac.es}
\author[orcid=0000-0002-4445-1845]{Matteo Pinamonti} 
\affiliation{INAF -- Osservatorio Astrofisico di Torino, Via Osservatorio 20, 10025 Pino Torinese (TO), Italy}
\email{matteo.pinamonti@inaf.it}
\author[orcid=0000-0002-8786-2572]{Monica Rainer} 
\affiliation{INAF -- Osservatorio Astronomico di Brera, via E. Bianchi 46, 23807 Merate (LC), Italy}
\email{monica.rainer@inaf.it}
\author[0009-0005-0463-1162]{Carolina Rodriguez} 
\affiliation{Observatoire astronomique de l'Universit\'e de Gen\`eve, Chemin Pegasi 51, 1290 Versoix, Switzerland}
\email{Carolina.Rodriguez@unige.ch}
\author[0000-0002-4829-7101]{Pam~Rowden} 
\affiliation{Royal Astronomical Society, Burlington House, Piccadilly, London W1J 0BQ, UK}
\email{prowden@ras.ac.uk}
\author[0000-0002-2394-0711]{Miquel Serra-Ricart} 
\affiliation{Instituto de Astrof\'isica de Canarias (IAC), Calle V\'ia L\'actea s/n, E-38205, La Laguna, Tenerife, Spain}
\affiliation{Departamento de Astrof\'isica, Universidad de La Laguna (ULL), E-38206 La Laguna, Tenerife, Spain}
\affiliation{Light Bridges, Observatorio Astron\'omico del Teide, Carretera del Observatorio s/n, E-38500 Guimar, Tenerife, Canarias, Spain}
\email{miquel@lightbridges.es}
\author[orcid=0000-0002-7504-365X]{Alessandro Sozzetti} 
\affiliation{INAF -- Osservatorio Astrofisico di Torino, Via Osservatorio 20, 10025 Pino Torinese (TO), Italy}
\email{alessandro.sozzetti@inaf.it}
\author[0000-0003-0647-6133]{Ivan A. Strakhov} 
\email{strakhov.ia15@physics.msu.ru}
\affiliation{Sternberg Astronomical Institute Lomonosov Moscow State University, 13 Universitetskij pr., 119234, Moscow, Russia}
\author[orcid=0000-0001-9261-8366, gname=Jhon, sname=Yana Galarza]{Jhon Yana Galarza} 
\email{jyanag@udec.cl}
\affiliation{Departamento de Astronomía, Facultad de Ciencias Físicas y Matemáticas Universidad de Concepción, Av. Esteban Iturra s/n Barrio Universitario, Casilla 160-C, Chile}
\author[orcid=0000-0001-7961-3907]{Samuel~W.~Yee} 
\email{syee@astro.ucla.edu}
\altaffiliation{51 Pegasi b Fellow}
\affiliation{Department of Physics \& Astronomy, University of California Los Angeles, Los Angeles, CA 90095, USA}

\begin{abstract}
Recent exoplanet demographics studies have uncovered the Neptune ridge: an overdensity of intermediate-size planets at orbital periods between 3--6~d. The origin of this pile-up remains unclear, but it could be a consequence of high-eccentricity migration (HEM). Here, we confirm TOI-4129\,b, a planet that may help to reveal the origin of the Neptune ridge. Using radial velocities from HARPS-N and transit photometry from TESS and four ground-based telescopes, we found that TOI-4129\,b has a relatively low density ($R_p = 5.7 \pm 0.2\, R_\oplus$, $M_p = 26\pm 6 \,M_\oplus$, $\rho_p = 0.8 \pm 0.2 ~\textrm{g}~\textrm{cm}^{-3}$) and orbits a $V = 12.9$ K dwarf star every 3.27~d, placing it at the inner edge of the Neptune ridge. We detect the stellar rotation period at 12.7~d in the photometry, implying a fairly young gyrochronal age of $1.20^{+0.16}_{-0.30}~$Gyr for the system. With a Transmission Spectroscopy Metric of $112\pm 30$, this adolescent Neptune is a good target for atmospheric observations with \textit{JWST} and \textit{Ariel}, and we also find that it is amenable to future constraints on the (three-dimensional) stellar obliquity. TOI-4129\,b will therefore be an important testing ground for the idea that planets in the Neptune ridge underwent HEM to arrive at their current locations.

\end{abstract}

\section{Introduction} \label{sec:Intro}

Neptune-sized exoplanets (${3R_\oplus} \leq R_p \leq 7R_\oplus$,as defined by \citealt{Naponiello_2025} for the HONEI program) are a rare but informative outcome of planet formation. Close-in ($P\lesssim100$~d) Neptunes can be divided into three demographic categories. First, the ``Neptune desert'' \citep{Latham2011, Szabo2011, 2016A&A...589A..75M} refers to the sparse population of hot Neptunes ($P \lesssim 3.2~\mathrm{d}$) that tend to be fairly dense \citep{Castro-Gonzalez2024b} and orbit metal-rich stars \citep[][L. Mancini et al. submitted]{2021AJ....162...62D, 2025AJ....169..117V, 2025A&A...701A.230M}. The more temperate Neptune savanna hosts longer-period \citep[$P \gtrsim 5.7~\mathrm{d}$;][]{Bourrier2023} planets with lower densities \citep{Castro-Gonzalez2024b}, exhibiting no clear host star metallicity preference \citep[][L. Mancini et al. submitted]{Dong2018, 2018AJ....155...89P, 2025AJ....169..117V, 2025A&A...701A.230M}. Nestled between the two, the ``Neptune ridge'' \citep[$3.2~\mathrm{d} \lesssim P \lesssim 5.7~\mathrm{d}$;][]{Castro_Gonz_lez_2024} is a recently-discovered overdensity of Neptunes reminiscent of the ``three-day pileup'' for hot Jupiters \citep{Dawson_2018}. Planets within the ridge possess modest densities \citep[typically $\rho < 2~\textrm{g}~\textrm{cm}^{-3}$;][]{Castro-Gonzalez2024b} and many possess excited eccentricities \citep{2020A&A...635A..37C} and misaligned orbits relative to the stellar spin axis \citep{Bourrier2023, 2025AJ....170..275Y, Bourrier2025}. Most of these ridge residents also seem to orbit metal-rich stars, similar to the desert dwellers \citep[][L. Mancini et al. submitted]{Dong2018, 2018AJ....155...89P, Petigura2022, 2025AJ....169..117V}. High-eccentricity migration \citep[HEM;][]{rasio1996, wu2003} has emerged as a possible formation mechanism for the dynamically-excited population of ridge planets \citep{Bourrier2018, Owen2018, Castro-Gonzalez2026}, although this mechanism may require finely-tuned stellar and planetary parameters for planets to avoid tidal disruption in the midst of migration \citep{2013ApJ...762...37L, Hallatt2026}.

However, our ability to test the HEM hypothesis remains limited, as many Neptune ridge planet candidates detected by the Transiting Exoplanet Survey Satellite \citep[TESS;][]{2015JATIS...1a4003R} remain unconfirmed. In particular, the confirmation of younger analogs to the mature Neptune population would be valuable, as measuring the primordial atmospheric compositions could be highly informative about the origins of Neptune-sized exoplanets. Additionally, HEM is predicted to result in misaligned orbits for Neptunes in the ridge \citep{Hallatt2026, Zanazzi2026}, so refining our knowledge of the obliquity distribution for these planets is a major active area of research \citep{Bourrier2023, Knudstrup2024, EspinozaRetamal2026, Mantovan2026, Wang2026}. This is a challenging observational endeavor. Neptunes are intrinsically uncommon \citep[relative to the more numerous sub-Neptunes; e.g.][]{Fulton2017}, and, moreover, they are inherently unfavorable for Rossiter-McLaughlin (RM) measurements relative to hot Jupiters due to their smaller radii. Therefore, efforts to confirm new Neptunes that are amenable to obliquity constraints (especially constraints on the three-dimensional obliquity $\psi$) are highly valuable.

In this paper, we report the confirmation of a new Neptune-sized exoplanet from TESS: TOI-4129\,b. This is our latest discovery from the Hot Neptune Initiative \citep[HONEI;][]{Naponiello_2025}. In Section~\ref{sec:obs}, we describe the TESS observations that revealed the initial transit signals and the accompanying photometric, spectroscopic, and imaging observations that aided in confirming the planet. In Section~\ref{sec:analysis}, we present the joint fit of RV and transit data using \textsf{jaxoplanet}. We continue to discuss the implications of the planet's density, age, and prospects for future atmospheric and obliquity constraints in Section~\ref{sec:discussion}. We conclude with a brief overview of our findings in Section~\ref{sec:conclusion}.

\section{Observations} \label{sec:obs}

\begin{figure}[t]
\includegraphics[width=\columnwidth]{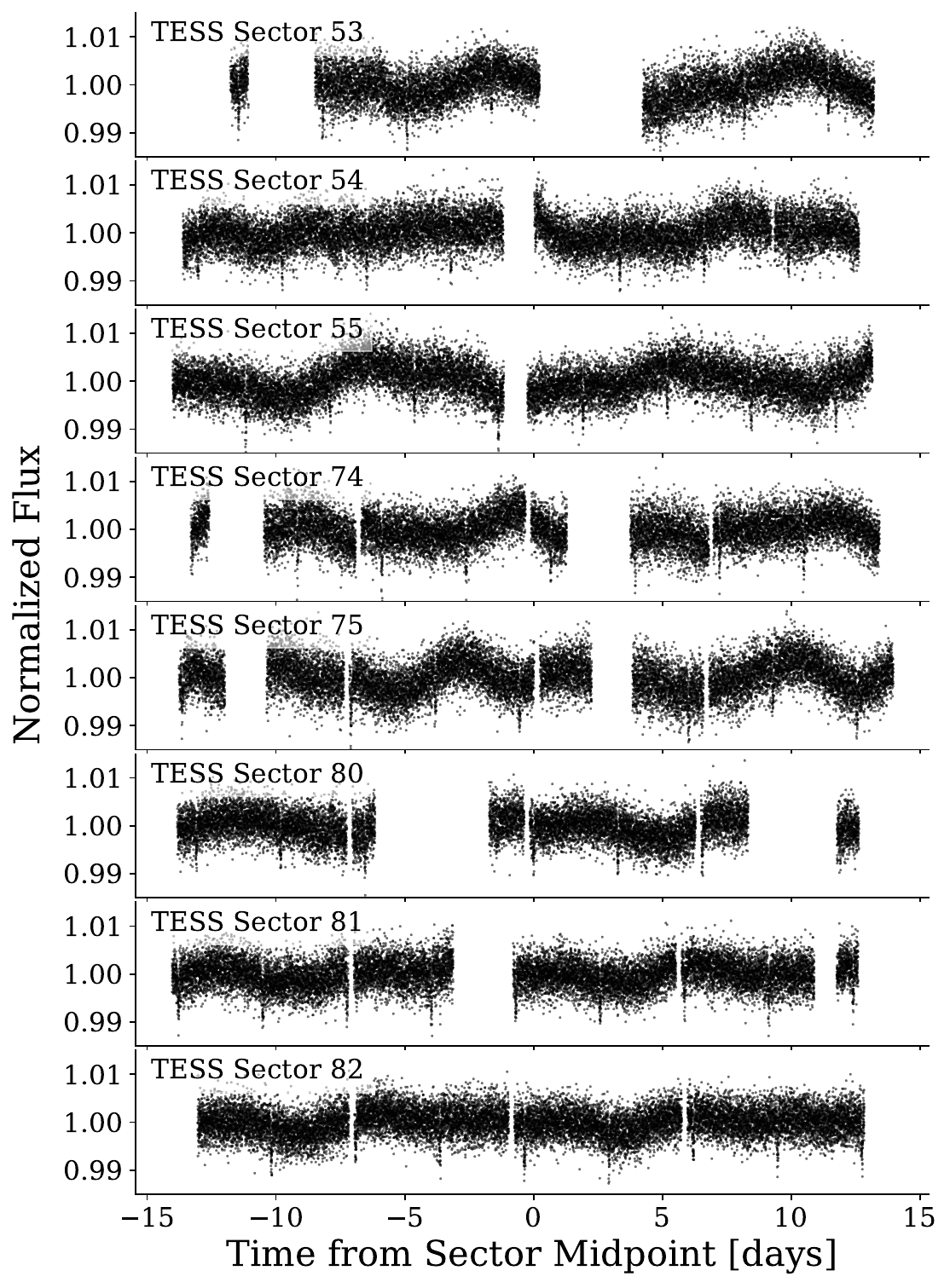}
\caption{PDCSAP light curves for all eight sectors of 120 s cadence TESS data for TOI-4129. The $12.7$~d stellar rotation period is clearly visible.  
\label{fig:TESS, wrapfig}}
\end{figure}

\subsection{TESS Photometry} \label{subsec:photometry}
TESS monitored the star TIC\,123357034\footnote{The target was almost observed by {\it Kepler} (KIC\,9568859) but ultimately fell outside the {\it Kepler} field.} from 2019 July to 2024 September. The star is relatively faint, with $V=12.9$; additional stellar parameters can be found in Table~\ref{tab:stellar}. The TESS Project designated the source as TOI-4129 in 2021 June as a result of the TESS faint-star search \citep{Kunimoto_2022}. TESS monitored this star with three different cadences: 1800~s (Sectors 14 and 26), 600~s (Sectors 40 and 41), and 120~s (Sectors 53, 54, 55, 74, 75, 80, 81, and 82). The star will be observed at 120~s cadence again in Sectors 118 and 120 (2027 June and 2027 August, respectively), which may provide a good opportunity for simultaneous photometry with future follow-up observations. We downloaded the PDCSAP light curves for TOI-4129 from the Mikulski Archive for Space Telescopes (MAST) using the \textsf{lightkurve} package \citep{2018ascl.soft12013L}. The PDCSAP 120~s cadence light curves are shown in Figure~\ref{fig:TESS, wrapfig}.

We first detrended the light curves to isolate the signal of the candidate planet from time-correlated astrophysical and instrumental signals. We temporarily masked transits using the period, transit epoch, and $1.5\times$ the duration initially reported by the TESS team, and then applied a Savitzky-Golay filter \citep{1964AnaCh..36.1627S} to determine the long-term trends in the photometry. We used different window lengths for each cadence (corresponding to a fixed window duration of roughly $5\times$ the transit duration): 225 for the 120~s cadence data, 45 for the 600~s cadence data, and 15 points for the 1800~s data. During this process, we also sigma-clipped 5$\sigma$ outliers above the median flux value (no data points were clipped below the median to avoid masking the transits). After dividing the light curves by the correlated noise model, the resulting light curves appeared to be free of systematics. As we were detrending the data, we noticed that the variability was quasiperiodic (which can be seen by eye in Figure~\ref{fig:TESS, wrapfig}), possibly as a consequence of the stellar rotation. We return to this signal in Section~\ref{subsec:age}.

\subsection{Ground-Based Photometry\label{sec:groundtransits}}

\begin{figure}[h!]
\plotone{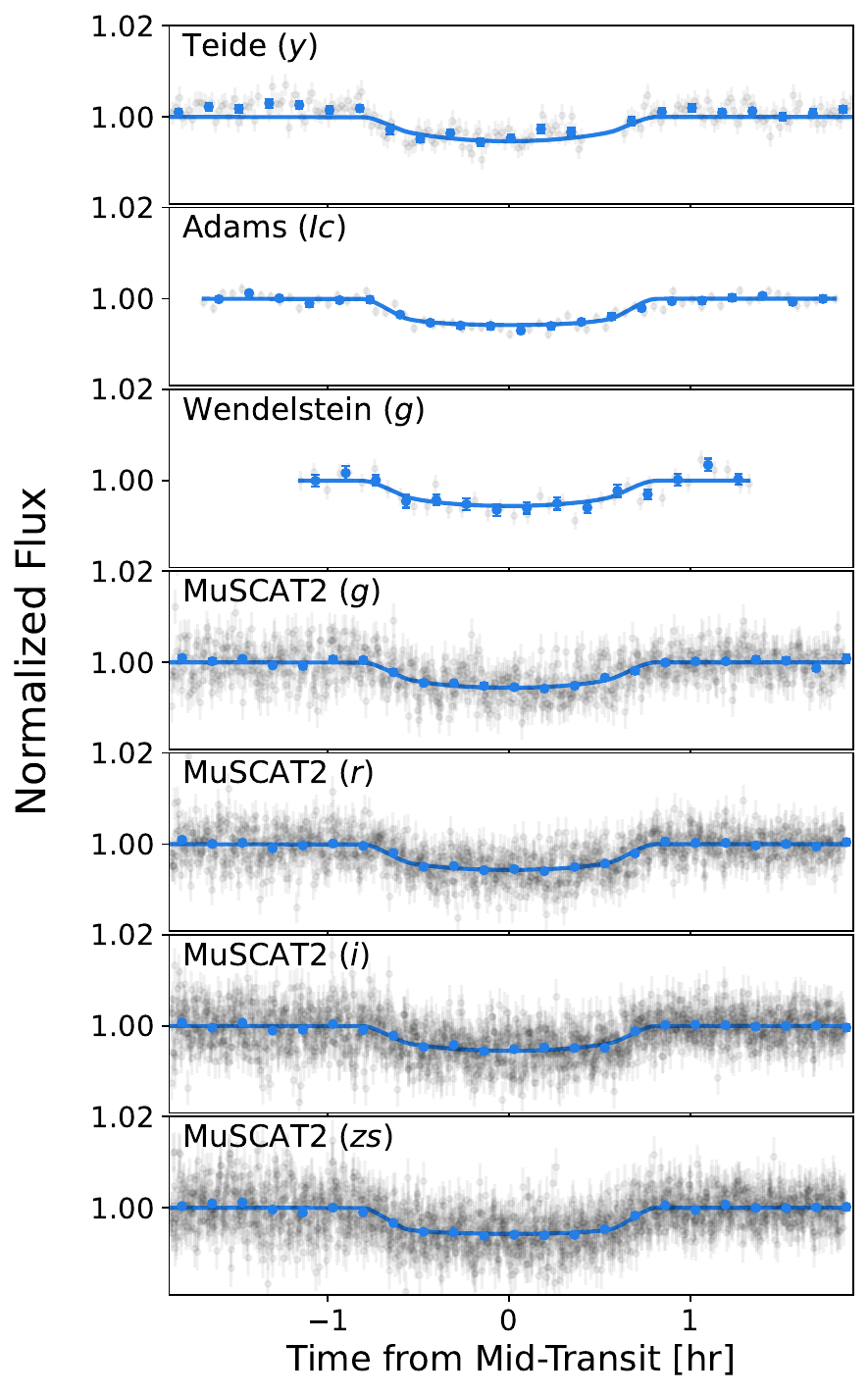}
\caption{Phase-folded light curves from ground-based datasets. The unbinned data are shown in gray and data are binned to 10 minute cadence in blue. The blue curves indicate our best-fit model (jointly fit with the TESS photometry and the HARPS-N RVs).  
\label{fig:stack}}
\end{figure}

The TESS pixel scale is $\sim 21\arcsec$ pixel$^{-1}$ and photometric apertures typically extend out to roughly $1\arcmin$, generally allowing multiple stars to blend in the TESS photometric aperture. To determine the true source of the TESS detections, we acquired ground-based time-series follow-up photometry of the field around TOI-4129 as part of the TESS Follow-up Observing Program Sub Group 1 \citep[TFOP SG1;][]{collins:2019}\footnote{https://tess.mit.edu/followup}. We used the {\tt TESS Transit Finder}, which is a customized version of the {\tt TAPIR} software package \citep{Jensen:2013}, to schedule our transit observations.

\subsubsection{Adams Observatory}

We observed a full transit of TOI-4129\,b in Johnson/Cousins $I$ band on UTC 2021 August 10 from the Adams Observatory at Austin College in Sherman, TX. The 0.6\,m telescope is equipped with an FLI ProLine detector that has an image scale of $0\farcs38$ pixel$^{-1}$, resulting in a $26\arcmin\times26\arcmin$ field of view. The differential photometric data were extracted with {\tt AstroImageJ} \citep{Collins:2017} using circular photometric apertures with radius $4\farcs2$. The target star aperture excluded flux from all Gaia DR2 neighbor stars, confirming that the transit is indeed occurring on-target. The final light curve is shown in Figure~\ref{fig:stack}. 

\subsubsection{MuSCAT2 Observations}

TOI-4129 was observed on UTC 2023 April 14 using the MuSCAT2 multi-band imager \citep{Narita2019} mounted on the 1.52~m Telescopio Carlos S\'{a}nchez (TCS) at Teide Observatory, Spain. MuSCAT2 is equipped with four CCDs capable of simultaneous imaging in the $g'$, $r'$, $i'$, and $z_s$ bands. Each $1024 \times 1024$ pixel CCD provides a field of view of $7.4 \times 7.4$ arcmin$^2$. The observations were conducted with the telescope in nominal focus. Exposure times were set to 15~s for the $g'$ band, 10~s for $r'$, and 7~s for the $i'$ and $z_s$ bands. Raw data were processed using the MuSCAT2 pipeline \citep{Parviainen2019}, which performs dark and flat-field calibrations, aperture photometry, and transit-model fitting including instrumental systematics. These observations unambiguously detected transits on the expected star, and demonstrated that the event was achromatic. The MuSCAT2 light curves are shown in Figure~\ref{fig:stack}. 

\subsubsection{Wendelstein Observatory}

From ExoFOP, we retrieved another publicly available full-transit light curve of TOI-4129\,b from the Wendelstein Observatory in Germany. The $0.4$~m telescope is equipped with an SBIG STX-16803 camera with an image scale of $0\farcs635$ per pixel. A full transit was observed in Sloan $g'$ on UTC 2022 June 03. {\tt AstroImageJ} \citep{Collins:2017} was used for image calibration and aperture photometry. Circular apertures with radius $5\farcs1$ were used for the aperture photometry, which excluded flux from all Gaia DR2 neighbor stars. The transit was clearly detected on-source, and is shown in Figure~\ref{fig:stack}.

\subsubsection{TTT Photometry}

A full transit of TOI-4129\,b was observed with the Two-meter Twin Telescope (TTT\footnote{\url{https://ttt.iac.es/}}), located at the Teide Observatory, Canary Islands in Spain. The facility consists of two identical 2.0-m f/6 Ritchey-Chrétien telescopes (TTT3/4) and two smaller replicas of 0.80-m f/6.85 (TTT1/2), all mounted on Alt-Az systems and equipped with multiple instruments at the Nasmyth foci.
The telescopes are equipped with the FERVOR (\textit{Fast Embedded-sCMOS Robotic Visible Observatory for Rapid transients}) detector, an optical imager based on sCMOS sensors (Sony IMX455 and IMX411; \citealt{Alarcon2023}).

Additionally, the TTT3 facility is equipped with the 2k$\times$2k  CCD Observatory for Light with Optical Robotic Six-band photometry (COLORS) mounted at the Nasmyth 2 focus, having a plate scale of 0\farcs23~px$^{-1}$ and a field of view of 7\farcm85 $\times$ 7\farcm85.
The TOI-4129 observation was carried out on UTC 2026 July 01 using COLORS@TTT3 in the SDSSy filter with an exposure time of 60~s.
The data were calibrated in the standard way using bias, dark and flat-field frames. The differential photometry was performed using a 3.5$\arcsec$ uncontaminated aperture using the {\tt AstroImageJ} software.

\subsection{High-Resolution Imaging}
As part of our standard process for validating transiting exoplanets to assess the possible contamination of bound or unbound companions on the derived planetary radii \citep{ciardi2015}, we observed TOI-4129 with near-infrared adaptive optics imaging on Keck and optical speckle imaging at SAI.  The near-infrared and optical imaging complement each other with differing resolutions and sensitivities.

Observations of TOI-4129 were made on UTC 2022 June 23 with the NIRC2 instrument on Keck-II (10~m) behind the natural guide star AO system \citep{wizinowich2000} in the standard 3-point dither pattern that is used with NIRC2 to avoid the left lower quadrant of the detector. The dither pattern with a $3\arcsec$ step size was performed three times, with each dither pattern offset from the previous dither by $0.5\arcsec$.  NIRC2 was used in the narrow-angle mode with a full field of view of $\sim10\arcsec$ and a pixel scale of approximately $0.0099442\arcsec$ per pixel. The Keck observations were made in the $K$ filter $(\lambda_o = 2.196; \Delta\lambda = 0.336~\mu$m). The integration times for an individual frame were 0.5~s for a total on-source time of 4.5~s. Flat fields were taken on-sky, dark-subtracted, and median averaged, and sky frames were generated from the median average of the dithered science frames. Each science image was then sky-subtracted and flat-fielded. The reduced science frames were combined into a single mosaiced image, with a final combined resolution of 0.052\arcsec.  
	
\begin{figure*}[b]
\includegraphics[width=\textwidth]{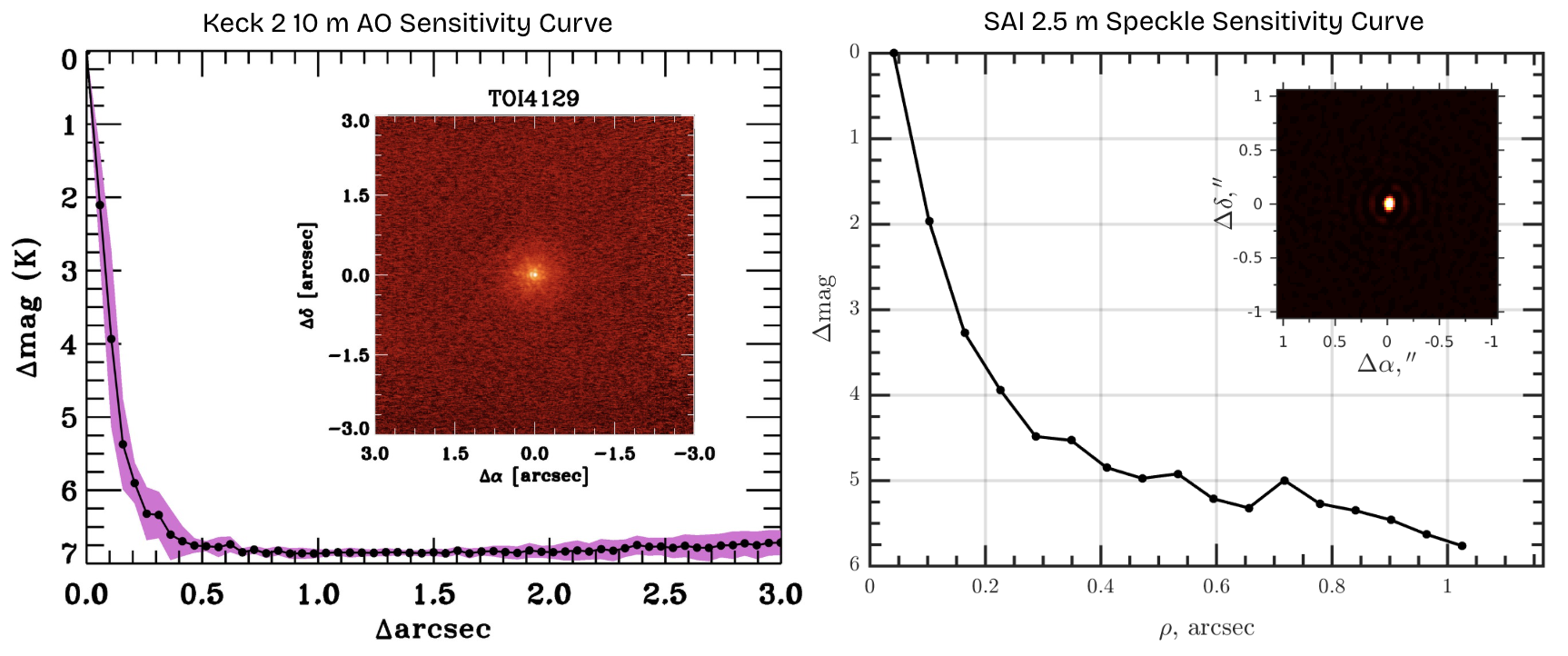}
\caption{Sensitivity curves from Keck/NIRC2 and the SAI speckle polarimeter. Keck/NIRC2 rules out companions with $\Delta K \lesssim 7$ past 0\farcs4, and SAI speckle polarimetry rules out companions with $\Delta I_c \lesssim 5$ past 0\farcs4. 
\label{fig:AO}}
\end{figure*}

The sensitivity of the final combined AO image was determined by injecting simulated sources azimuthally around the primary target every $20^\circ $ at separations of integer multiples of the central source's FWHM \citep{furlan2017}. The brightness of each injected source was scaled until standard aperture photometry detected it with $5\sigma$ significance.  The final $5\sigma$ limit at each separation was determined from the average of all of the determined limits at that separation and the uncertainty on the limit was set by the rms dispersion of the azimuthal slices at a given radial distance; sensitivities are shown in Figure~\ref{fig:AO}.

TOI-4129 was also observed on July 20, 2021 with the speckle polarimeter on the 2.5-m telescope at the Caucasian Observatory of Sternberg Astronomical Institute (SAI) of Lomonosov Moscow State University \citep{AL190145}. An electron multiplying CCD Andor iXon 897 was used as a detector. The atmospheric dispersion compensator was active, which allowed the use of the $I_\mathrm{c}$ band. The angular resolution was 0\farcs083. Long exposure seeing, as estimated from the data, was moderate, 0\farcs91. No companions were detected. The detection limits at separations of 0\farcs25 and 1\farcs0 are $\Delta I_\mathrm{c}=4.2$ and $5.7$, respectively.

Both imaging observations rule out a bright secondary star farther than $0\farcs4$ from the host star, with the images reaching a contrast of $\approx$ 5 (SAI) and $\approx$ 7 (Keck-II) magnitudes fainter than TOI-4129 within each respective band.

\subsection{Radial Velocities} \label{subsec:RV}

\subsubsection{TRES Spectra}

We obtained 18 spectra from the Tillinghast Reflector Echelle Spectrograph (TRES), which is an $R\sim44,000$ optical echelle spectrograph on the 1.5 meter Tillinghast telescope at the Fred Lawrence Whipple Observatory \citep{2007RMxAC..28..129S}. The data were collected between 2021 July 13 and 2024 May 24 with exposure times ranging from 1300 s to 3600 s and an average internal RV uncertainty of 23.6~m~s$^{-1}$. The phased RV curve is shown in Figure~\ref{fig:TRES, wrapfig}. While not precise enough to confirm the planet, the TRES data motivated us to follow up the target with more precise RV observations.

\begin{figure}[t]
\includegraphics[width=\columnwidth]{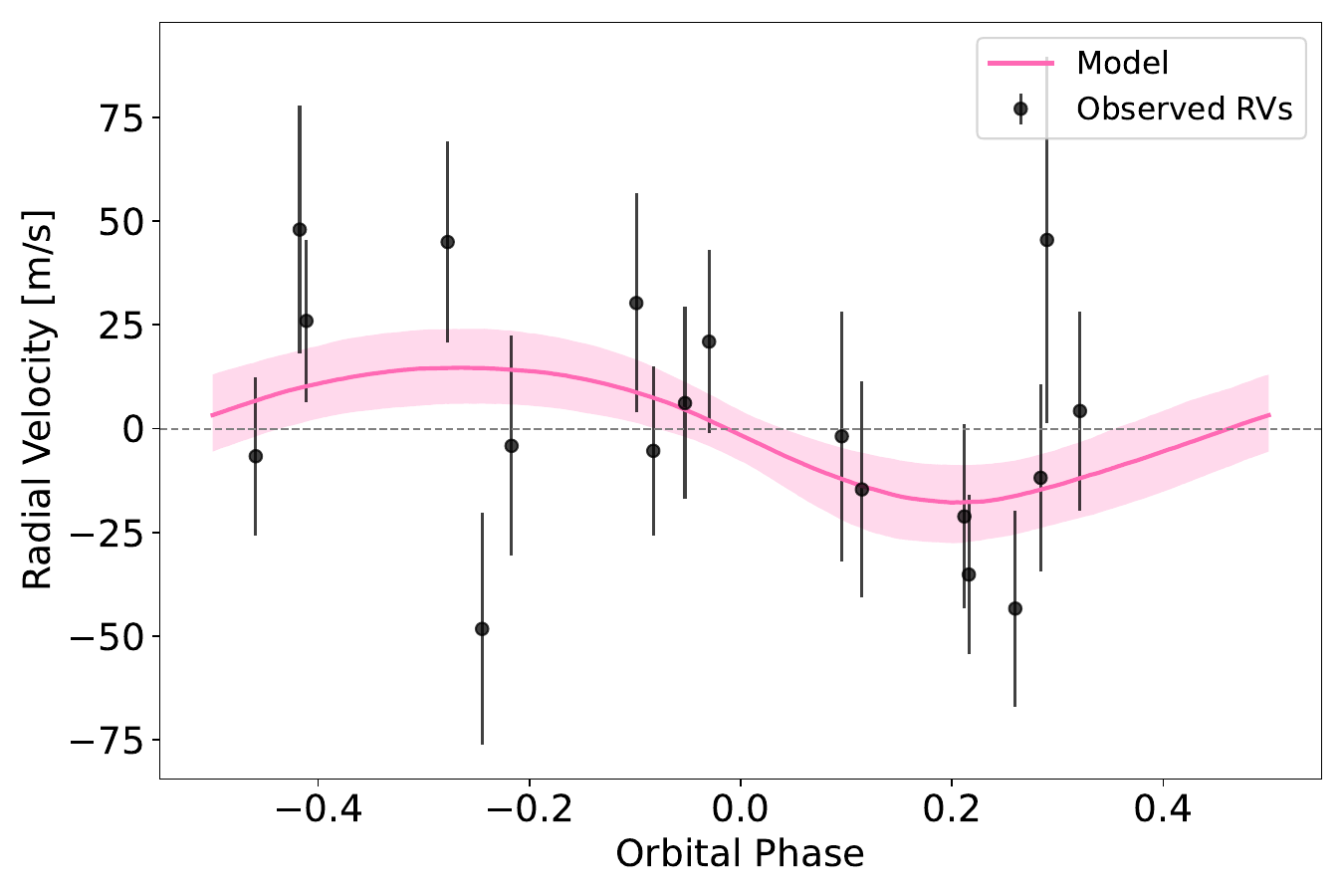}
\caption{RV measurements from TRES phased to the planet's orbital period. The best-fit model from a fit to TRES RV and TESS transit data is shown in pink, with the shaded region denoting the $1\sigma$ uncertainty interval. While the reflex motion of the star is not clearly detected with the TRES data alone, an eclipsing binary is ruled out.
\label{fig:TRES, wrapfig}}
\end{figure}

\subsubsection{HARPS-N Spectra}

We obtained 25 epochs of precise RV observations from the High Accuracy Radial velocity Planet Searcher for the Northern Hemisphere \citep[HARPS-N;][]{2012SPIE.8446E..1VC}, a high-resolution echelle spectrograph located on the 3.58 m Telescopio Nazionale Galileo. The measurements were taken between 2023 October 9 and 2024 June 23. The data were processed with the HARPS-N Data Reduction Software \citep[DRS;][]{Dumusque_2021} version 3.3.12. The spectra have a median S/N of 24.6 in order 50, and a K6 stellar mask was used for determining the RVs via the cross-correlation function method. We found the star to be fairly active, with Mt. Wilson index $S_\mathrm{MW} = 1.12\pm0.10$. This corresponds to $\log(R_\mathrm{HK}') = -4.33\pm0.08$ using the \citet{Astudillo-Defru2017} calibration and the $V-K$ color calculated from our Table~\ref{tab:stellar}. The average internal RV uncertainty was 5.5~m~s$^{-1}$, roughly as expected for the relatively faint $V=12.9$ star. This was sufficient to clearly detect the reflex motion of the star.

\section{Data Analysis} \label{sec:analysis}

\subsection{Stellar Characterization\label{subsec:stellar}}

\begin{deluxetable*}{rlll}[p]
\digitalasset
\tablewidth{0pt}
\tablecaption{Stellar Parameters for TOI-4129 \label{tab:stellar}}
\tablehead{
\colhead{Parameter} & \colhead{Value} & \colhead{Units} & \colhead{Source}
}
\startdata
Cross-Identifiers & {} & {} & {} \\
TOI & 4129 & & TOI \\ 
TIC &  123357034 & & TIC \\
2MASS & J18460652+4614546  & & 2MASS \\
Gaia DR3 & 2119155349385688576 &  & Gaia DR3 \\
KIC & 9568859 &  & KIC \\
\hline
Astrometry & {} & {} & {} \\
$\alpha$ (J2016)  & 281.52727768779 & ° & Gaia DR3 \\
$\delta$ (J2016) & +46.24856779675 & ° & Gaia DR3 \\
$\pi$  & $7.64\pm0.01$ & mas &  Gaia DR3 \\
$\mu_{\alpha}\cos{\delta}$  & $10.795\pm0.012$ & mas yr$^{-1}$ & Gaia DR3 \\
$\mu_\delta$  & $12.591\pm0.013$ & mas yr$^{-1}$ & Gaia DR3 \\
\hline
Photometry & {} & {} & {} \\
TESS  & $11.8417\pm0.0061$ &  & TIC \\
\textit{Gaia} & $12.5596\pm0.0028$ &  & Gaia DR3 \\
$V$  & $12.895\pm0.103$ &  & TIC \\
$J$  & $10.801\pm0.021$ &  & 2MASS \\
$H$  & $10.232\pm0.021$ &  & 2MASS \\
$K_s$  & $10.092\pm0.011$ &  & 2MASS \\
\textit{WISE}~3.4$~\mu$m  & $10.032\pm0.023$ &  & AllWISE \\
\textit{WISE}~4.6$~\mu$m   &  $10.100\pm0.021$ &  & AllWISE \\
\textit{WISE}~12$~\mu$m  & $10.008\pm0.04$ &  & AllWISE \\
\textit{WISE}~22$~\mu$m  & $8.976\pm0.326$ &  & AllWISE \\
\hline 
Spectroscopic Parameters & {} & {} & {} \\
Spectral Type & K & & This work \\
$[\mathrm{Fe}/\mathrm{H}]$ & $-0.04\pm0.09$ & dex & This work \\
\vsini & $<3$ & km s$^{-1}$ & This work \\
\hline
Derived Parameters & {} & {} & {} \\
$T_\textrm{eff}$ & $4564\pm63$ & K & This work \\
$M_\star$  & $0.698\pm 0.028$ & $M_\odot$ & This work\\
$R_\star$ & $0.671\pm0.018$ & $R_\odot$ & This work \\
$L_\star$ & $0.176\pm0.011$ & $L_\odot$ & This work \\
$\rho_{\star}$ & $3.25\pm0.28$ & $\textrm{g}~\textrm{cm}^{-3}$  & This work \\
$\log(g)$ & $4.628\pm0.025$ & cgs & This work \\
Distance & $130.85\pm0.18$ & pc & This work \\
$S_\mathrm{MW}$ & $1.12\pm0.10$ & & This work \\
Isochronal Age & 6.5$^{+4.8}_{-4.3}$~ & Gyr &  This work \\
Gyrochronal Age & 1.20$^{+0.16}_{-0.30}$~ & Gyr &  This work \\
$P_{\mathrm{rot}}$ & 12.72$~\pm~0.17$~ & d &  This work \\
\enddata
\tablerefs{TESS Primary Mission TOI catalog: \citealt{Guerrero_2021}; TIC v8.2: \citealt{2022yCat.4039....0P}; Gaia DR3: \citealt{refId0}; 2MASS: \citealt{2003yCat.2246....0C}; AllWISE: \citealt{2014yCat.2328....0C}}
\end{deluxetable*}

The stellar atmospheric parameters of TOI-4129, namely the effective temperature \teff, surface gravity \logg, microturbulence velocity \csi, and iron abundance \feh, were derived from the co-added HARPS-N spectrum of the target. This spectrum was obtained by co-adding all the collected spectra, resulting in a final per-pixel signal-to-noise ratio of 170 at $\sim$6000\,\AA. We followed the procedure described in \citet{Biazzo_2022} to determine the stellar parameters using the equivalent-width (EW) method based on \ion{Fe}{1} and \ion{Fe}{2} lines. The line equivalent widths were measured with IRAF. Lines with $\mathrm{EW}>150$ m\AA\, and those with uncertainties larger than $2\sigma$ were discarded from the analysis. We adopted the line list presented in \citet{Biazzo_2022} and employed 1D model atmospheres linearly interpolated from the ATLAS9 grid of \citet{castelli2004newgridsatlas9model}, assuming a solar-scaled chemical composition and updated opacities (ODFNEW). The final stellar parameters and iron abundance were derived using the pyMOOGi code \citep{2017AAS...23021607A}, a python wrapper of the {\sc MOOG} code (\citealt{osti_4288566}, version 2019), employing the abfind driver and assuming local thermodynamic equilibrium (LTE). They were obtained by iteratively changing \teff, \csi, and \logg\ at steps of 5\,K, 0.01\,km~s$^{-1}$, and 0.01\,dex, respectively, until the slopes of the line abundances with the reduced equivalent width $\mathrm{REW}=\log(\mathrm{EW}/\lambda)$ and the excitation potential were close to zero, and until the \ion{Fe}{1} abundance was close to the \ion{Fe}{2} abundance (see \citealt{Biazzo_2022} for details). Comparing the derived \teff\,with the v.2022 calibration table of \citet{2013ApJS..208....9P}, we find that TOI-4129 is consistent with a K4 dwarf star. The results of the spectroscopic analysis are reported in Table\,\ref{tab:stellar}. We also report the projected rotational velocity ($v\sin i$) of TOI-4129 using the calibration between the FWHM of the CCF of the co-added spectrum and the ($v\sin{i}$) established by \citet{Rainer2023}. However, we caution that this value should be treated as an upper limit, as this relation is not well-calibrated in the regime of low $v\sin i$.

Then, we determined the stellar mass, radius, and age through simultaneous modeling of the stellar Spectral Energy Distribution (SED) and the MIST evolutionary tracks \citep{Paxton2015, Choi2016, Dotter2016} in a Bayesian framework with the {\tt EXOFASTv2} tool \citep{2017ascl.soft10003E, Eastman2019}; see also \citet{naponiello2025a} for more details. To sample the SED, we used the 2MASS near-infrared $J$, $H$, and $K_s$ magnitudes and the WISE $W1$, $W2$, $W3$ infrared magnitudes (see Table~\ref{tab:stellar}). We adopted Gaussian priors on the Gaia DR3 parallax as well as on the $T_{\rm eff}$ and [Fe/H] derived from the analysis of the HARPS-N coadded spectrum. We then computed the values and $1\sigma$ uncertainties of the stellar mass, radius, and  age from the medians and 16\%\,$-$\,84\% percentiles of their posterior distributions: $M_\star=0.698\pm0.028\,M_\odot$, $R_\star=0.671\pm0.018\, R_\odot$, and  $t_{\rm age} = 6.5^{+4.8}_{-4.3}$~Gyr (see Table~\ref{tab:stellar}). As expected for a late K dwarf, the age from the MIST evolutionary tracks is largely unconstrained. By imposing a prior on the stellar age from gyrochronology (refer to Section~\ref{subsec:age}), we derived a slightly higher $M_\star$ and slightly lower $R_\star$, although these values are consistent within 1$\sigma$ with the solution we adopt here.

\subsection{Planet Characterization} \label{subsec:jointfit}

We jointly fit the transit and RV data using the \textsf{jaxoplanet} package \citep{jaxoplanet}. We included the ground-based light curves, the TESS light curves (treated separately for each cadence of data), and the HARPS-N RVs; we omitted the TRES RVs from the fit due to their lower precision compared to the HARPS-N dataset. We parameterized the model using the orbital period $P$, mid-transit time $t_0$, impact parameter $b$, planet radius $R_p$, planet mass $M_p$, stellar mass $M_\star$, stellar radius $R_\star$, the quadratic limb-darkening parameters $u_0$ and $u_1$, and the eccentricity parameters $\sqrt{e}\cos\omega$ and $\sqrt{e}\sin\omega$. We included error scaling terms $s_i$ for each of the light curves as well as jitter and offset terms for the HARPS-N RV dataset. The prior probability distributions for these model parameters are listed in Table~\ref{tab:orbital}. We chose to fit in this basis (rather than fitting for e.g. the RV semiamplitude $K$ or radius ratio $R_p/R_\star$) so we could set more physically-motivated priors on planetary parameters. To facilitate comparison with other works, we derive $T_{14}$, $R_p/R_\star$, $M_p/M_\star$, $a/R_\star$, $i$, $a$, $e$, $\omega$, $\rho_p$, $K_\star$, and $T_\mathrm{eq}$ in our fits and report them in Table~\ref{tab:orbital} as well.

\begin{figure*}[ht!]
\plotone{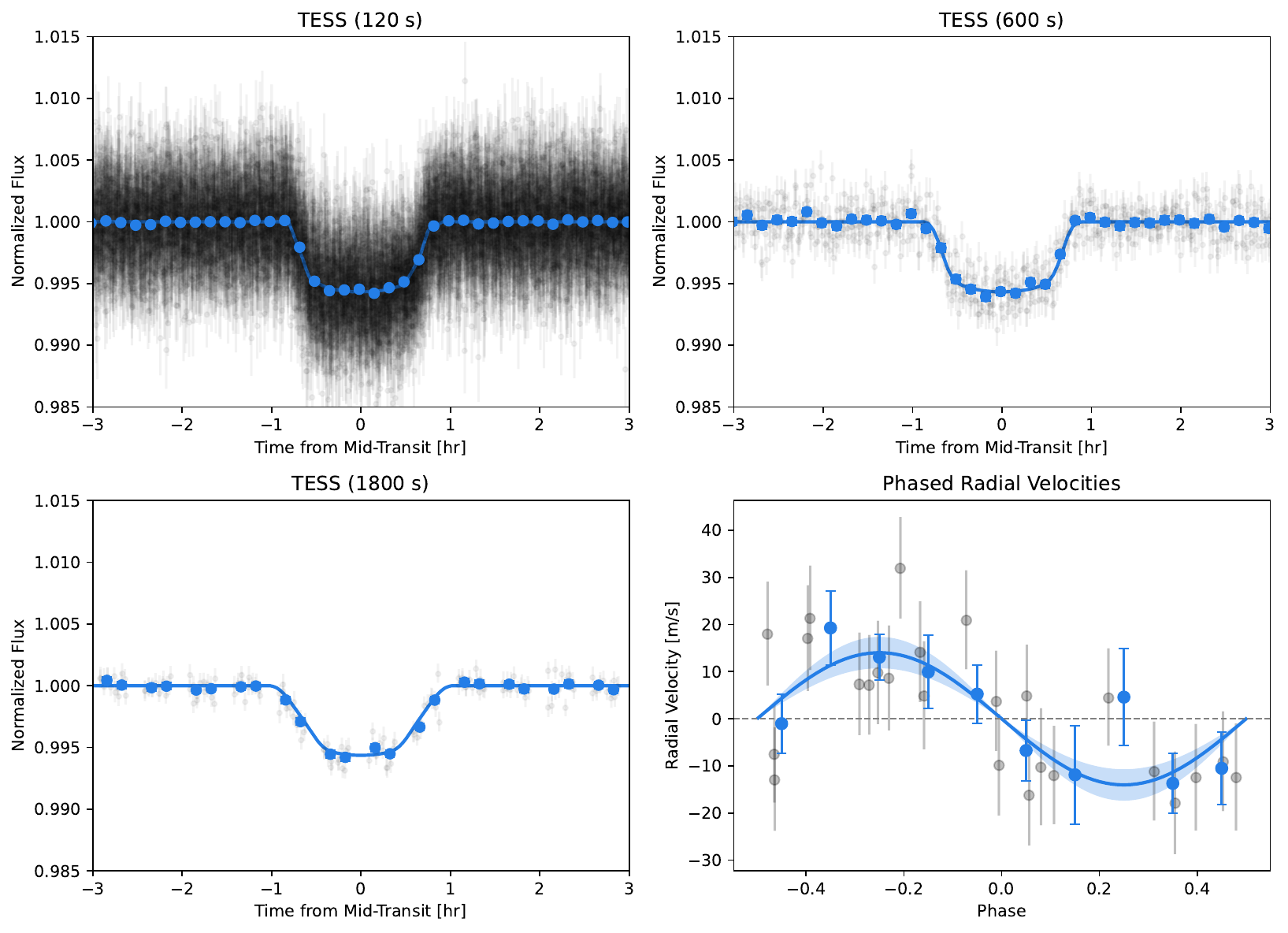}
\caption{Phase-folded TESS light curves and HARPS-N RVs (both in gray) with the best-fit circular ($e$ fixed) model given by the blue curves along with the associated shaded uncertainty intervals. The blue points show the light curves binned to 10~min cadence and the RVs in phase bins of width 0.1.
\label{fig:2x2}}
\end{figure*}

We sampled the posterior probability distributions for the model parameters using \textsf{numpyro} \citep{bingham2019pyro, phan2019composable}. We first used the optimization tools in \textsf{numpyro\_ext} to search for the maximum a posteriori (MAP) solution. We subsequently initialized our No U-Turn Sampler \citep{Hoffman2011} near the MAP solution, and then ran two chains for the fit, where each chain tuned for 1000 steps before 2000 posterior draws were taken. We verified that the Gelman-Rubin statistic \citep{1992StaSc...7..457G} was $\hat{R}\ll 1.01$ for each sampled parameter. The final light curves and the phased RVs are shown in Figure~\ref{fig:2x2} alongside the best-fit model and uncertainty intervals. The posterior probability distributions for all model parameters are summarized in Table~\ref{tab:orbital}. 

Given that our eccentricity constraint was weak ($e<0.20$ at 95\%), we ran a fit with the eccentricity fixed to zero, which resulted in posteriors within $1\sigma$ of the $e$-free posteriors. We assessed the impact of fixing the eccentricity using Bayesian model comparison, computing and comparing the leave-one-out (LOO) cross-validation score \citep{Vehtari_2016} for both models. The two models are statistically indistinguishable (the eccentricity-fixed model was preferred with an improved expected log pointwise predictive density of $\Delta\mathrm{ELPD_{LOO}}$ = 0.48), so we adopted the simpler fixed eccentricity model throughout the paper. The posteriors for the $e=0$ fit are also provided in Table~\ref{tab:orbital}.

\subsection{Non-Detection of Transit Timing Variations}
To investigate the presence of potential additional bodies in the system, we performed a Transit Timing Variation (TTV) analysis. We derived the individual mid-transit times by fitting each available transit epoch independently. The results do not exhibit any significant deviation from a linear ephemeris, yielding a weighted mean TTV of $0.16 \pm 0.30$~min. We further computed a Generalized Lomb-Scargle \citep[GLS;][]{Zechmeister2009} periodogram, revealing no significant periodic signals, with the highest peak remaining well below the $10\%$ False Alarm Probability (FAP) threshold. We therefore conclude that there is currently no evidence of dynamically significant TTVs for this target.

\begin{deluxetable*}{rllll}[p]
\digitalasset
\tablewidth{0pt}
\tablecaption{Orbital Parameters for TOI-4129\,b \label{tab:orbital}}
\tablehead{
\colhead{Parameter} & \colhead{Prior} & \colhead{Posterior ($e$ free)} & \colhead{Posterior ($e$ fixed)} & \colhead{Units}
}
\startdata
Fit Parameters & {} & {} & {} & {} \\
$P_\mathrm{orb}$  & $\mathcal{U}(3.27246,3.27253)$ & 3.2725003 $\pm~0.0000010$ & 3.27250031 $\pm~0.00000098$ & d \\
$T_0$ & $\mathcal{U}(2459744.23,2459744.39)$ & $2459744.30855\pm 0.00024$ & $2459744.30846\pm0.00016$  & BJD \\
$b$ & $\mathcal{U}(0.0,1.5)$ & $0.772\pm0.022$ & $0.755\pm0.016$ & -- \\
$R_p$ & $\mathcal{U}(0.0,10.0)$ & $5.60\pm 0.18$ & $5.67\pm 0.17$ & $R_\oplus$\\
$M_p$ & $\mathcal{U}(0.0,100.0)$ & $25.4\pm6.3$ & $25.5\pm6.3$ & $M_\oplus$ \\
$\sqrt{e}\cos\omega$ & unit disk & $0.10\pm0.19$ & & -- \\
$\sqrt{e}\sin\omega$ & unit disk & $-0.17\pm0.13$ & & -- \\
$M_\star$ & $\mathcal{N}(0.7,0.03)$ & $0.697\pm 0.030$ & $0.688\pm 0.029$ & $M_\odot$ \\
$R_\star$ & $\mathcal{N}(0.671, 0.02)$ & $0.677\pm0.020$ & $0.689\pm0.016$ & $R_\odot$ \\
\hline
Non-fitted Parameters & {} & {} & {} & {} \\
$T_{14}$ &  & 1.626 $\pm$ 0.019 & 1.614 $\pm$ 0.016 & hr \\
$R_p/R_\star$ &  & $0.07583 \pm 0.00099$ & $0.07535 \pm 0.00084$ & -- \\
$M_p/M_\star$ &  & $0.000109 \pm 0.000027$ & $0.000111 \pm 0.000027$ & -- \\
$a/R_\star$ &  & $12.16\pm0.39$ & $11.88\pm0.28$ & -- \\
$i$ &  &  $86.52 \pm0.22 $ &$86.35 \pm0.16 $ & ° \\
$a$ &  &  $0.03824\pm0.00054$ & $0.03807\pm0.00053$ & au \\
$ \rho_p$ &  & $0.80\pm 0.21$ & $0.78\pm 0.20$ & $\textrm{g}~\textrm{cm}^{-3}$ \\
$e$ &  &  $0.091 \pm 0.059$ & & -- \\
$\omega$ & & $-53 \pm 63$ & & ° \\
$K_\star$ &  & $13.9\pm3.4$ & $14.1\pm3.4$ & $\textrm{m}~\textrm{s}^{-1}$ \\
$T_\mathrm{eq}$ &  & $926\pm15$ & $936\pm11$ & K \\
\hline
Instrumental Parameters & {} & {} & {} \\
$s_{\mathrm{TESS},~2}$ & $\mathcal{U}(0.5, 5.0)$ & $0.9528\pm0.0019$ & $0.9528\pm0.0019$ & -- \\
$s_{\mathrm{TESS},~10}$ & $\mathcal{U}(0.5, 5.0)$ & $0.9532\pm0.0078$ & $0.9532\pm0.0078$ & -- \\
$s_{\mathrm{TESS},~30}$ & $\mathcal{U}(0.5, 5.0)$ & $0.6272\pm0.0096$ & $0.6273\pm0.0095$ & -- \\
$s_{\mathrm{Adams},~Ic}$ & $\mathcal{U}(0.5, 5.0)$ & $0.923\pm0.081$ & $0.928\pm0.083$ & -- \\
$s_{\mathrm{MuSCAT},~g}$ & $\mathcal{U}(0.5, 5.0)$ & $0.965\pm0.024$ & $0.965\pm0.023$ & -- \\
$s_{\mathrm{MuSCAT},~i}$ & $\mathcal{U}(0.5, 5.0)$ & $1.017\pm0.018$ & $1.017\pm0.018$ & -- \\
$s_{\mathrm{MuSCAT},~r}$ & $\mathcal{U}(0.5, 5.0)$ & $0.992\pm0.020$ & $0.992\pm0.020$ & -- \\
$s_{\mathrm{MuSCAT},~zs}$ & $\mathcal{U}(0.5, 5.0)$ & $1.008\pm0.017$ & $1.008\pm0.018$ & -- \\
$s_{\mathrm{Teide},~y}$ & $\mathcal{U}(0.5, 5.0)$ & $1.506\pm0.073$ & $1.508\pm0.076$ & -- \\
$s_{\mathrm{Wendelstein},~g}$ & $\mathcal{U}(0.5, 5.0)$ & $2.80\pm0.35$ & $2.80\pm0.35$ & -- \\
RV Jitter & $\mathcal{U}(0.0,20.0)$ & $9.5\pm1.9$ & $9.6\pm2.0$ & $\textrm{m}~\textrm{s}^{-1}$ \\
RV Offset & $\mathcal{U}(-26224.0, -26184.0)$ & $-26208.5\pm2.3$ & $-26208.6\pm2.3$ & $\textrm{m}~\textrm{s}^{-1}$ \\
\hline
Limb-Darkening Coefficients & {} & {} & {} & {} \\
$u_0$ (TESS) &  & $0.36\pm0.20$ & $0.34\pm0.20$ & -- \\
$u_1$ (TESS) &  & $0.14\pm0.25$ & $0.18\pm0.25$ & -- \\
$u_0$ (g) &  & $0.78\pm0.31$ & $0.85\pm0.29$ & -- \\
$u_1$ (g) &  & $0.04\pm0.36$ & $-0.03\pm0.34$ & -- \\
$u_0$ (i) &  & $0.99\pm0.23$ & $1.02\pm0.22$ & -- \\
$u_1$ (i) &  & $-0.27\pm0.27$ & $-0.30\pm0.26$ & -- \\
$u_0$ (Ic) &  & $0.28\pm0.21$ & $0.29\pm0.23$& -- \\
$u_1$ (Ic) &  & $0.35\pm0.30$ & $0.37\pm0.31$& -- \\
$u_0$ (r) &  & $0.60\pm0.29$ & $0.64\pm0.29$ & -- \\
$u_1$ (r) &  & $0.16\pm0.36$ & $0.12\pm0.35$ & -- \\
$u_0$ (y) & & $1.30\pm0.27$ & $1.29\pm0.29$& -- \\
$u_1$ (y) & & $-0.50\pm0.24$ & $-0.50\pm0.25$& -- \\
$u_0$ (zs) & & $0.23\pm0.18$ & $0.24\pm0.18$& -- \\
$u_1$ (zs) & & $0.25\pm0.25$ & $0.28\pm0.27$& -- \\
\hline
\enddata
\tablecomments{$T_\mathrm{eq}$ was calculated assuming a Bond albedo $A_\mathrm{B} = 0$, i.e. $T_\mathrm{eq} = T_\mathrm{eff}\sqrt{\frac{R_\star}{2a}}$.}
\end{deluxetable*}

\section{Discussion} \label{sec:discussion}

Here we examine the age of the TOI-4129 system, place planet b into context of the Neptune ridge population, and describe the planet's potential for future characterization.

\subsection{Gyrochronology} \label{subsec:age}

While performing the light curve detrending, we constructed an initial Lomb-Scargle periodogram \citep[LS;][]{1976Ap&SS..39..447L, 1982ApJ...263..835S} on the PDCSAP light curves with the transits masked out, and found that TOI-4129 has a stellar rotation period of approximately $12.72 \pm 0.17$~d (with the error bar representing the full-width at half-maximum for the LS peak). We repeated this experiment using the SAP light curve and found a consistent result. Based on this initial test, we sought out additional verifications. The TESS All-Sky Rotation Survey \citep[TARS;][]{boyle2026tessallskyrotationsurvey} team adopted a period of 12.49~d for TOI-4129. Gaia DR3 additionally previously noted TOI-4129 as variable \citep{refId0}, as did the Zwicky Transient Facility \citep[ZTF;][]{Bellm2019, Masci2019}. In particular, the strongest peak in the LS periodogram of the Gaia DR3 photometry has a FAP below 0.1\% and is located at 12.56~d. There are also consistent peaks in the ZTF periodograms for both the $g$ and $r$ band photometry. 

Given that multiple groups have determined similar rotation periods independently with different instruments, it is unlikely that this is a result of our detrending methodology. The signal is astrophysical rather than instrumental in nature. We attribute the quasiperiodic signal to stellar rotation, and use gyrochronology to estimate TOI-4129's age given its relatively rapid rotation period. With our measured stellar rotation period ($P_\mathrm{rot} = 12.7$~d) and effective temperature $T_\mathrm{eff}$ (Table~\ref{tab:stellar}), we used the \textsf{gyro-interp} package \citep{Bouma_2023} to determine a system age of 1.20$^{+0.16}_{-0.30}$~Gyr. The asymmetric age posterior indicates that TOI-4129 may be in a period of stalled spin-down experienced by late K dwarfs \citep{Curtis_2019, Curtis2020}. 

Motivated by the rotation signal and the relatively high jitter, we experimented with including a Gaussian process (GP) into our fitting of the RVs. We attempted a multi-GP fit using \textsf{pyaneti} \citep{Barragan2019, Barragan2022}, but found that none of the more complex models that we tested (including trends, additional planets, correlated noise from stellar activity, or free eccentricity) proved more significant than the simple circular one-planet model. According to the Multidimensional Gaussian Process Information Criterion \citep[MGIC;][]{Barragan2026} metric, the one-planet model with zero eccentricity and no activity indicators (as adopted in Table~\ref{tab:orbital}) performs better than the GP model by a margin of approximately $-10$. While the high jitter suggests there could be a second planetary signal hidden in the noise, we cannot prove it at this stage. More RVs will be needed to test this hypothesis.

As a secondary validation of our age estimate, we also consider the star’s galactic kinematics using astrometric information from Gaia DR3 \citep{refId0}. We find UVW space motions of ($-4, -7, -5$)\,km\,s$^{-1}$ relative to the local standard of rest \citep[LSR;][]{Schonrich2010}, resulting in a total space velocity of 9\,km\,s$^{-1}$. To contextualize this value, we note that \citet{Santos2003} found an average total space velocity of 42\,km\,s$^{-1}$ for planet-hosting stars, with a dispersion of 28\,km\,s$^{-1}$. The relatively low space velocity of TOI-4129 may suggest youth: at the population level, young stars are dynamically colder than old stars. While galactic kinematics cannot provide a precise age estimate for any individual star, we can consider whether the motion of TOI-4129 is statistically likely given the gyrochronologically determined age. Using the age–velocity-dispersion relation from \citet{Yu2018} and reparameterized in \citet{Lu2021}, stars at an age of 1.2~Gyr have a characteristic velocity dispersion of $\sigma_{v_z}=7$\,km\,s$^{-1}$, where $v_z$ is the velocity in the direction of the Galactic north pole. Note that $v_z=W_{\rm LSR}$, making $v_z=-5$\,km\,s$^{-1}$ in this case. Thus, the galactic kinematics of TOI-4129 are indeed plausibly consistent with an age of 1.2~Gyr.

We also searched for stellar Li\,I 6708~\AA~absorption as an extra check on the age estimate. We tentatively detected the presence of weak lithium absorption in the co-added stellar spectrum, with an equivalent width of EW\,=$2.5\pm1.0$\,m\AA. Adopting the spectroscopic parameters measured in this work, we determined an upper limit on the NLTE lithium abundance of A(Li)$<0.0$ dex \citep{Lind2009}. Based on empirical isochrones by \citet{Jeffries2023} and using the EW and $T_{\rm eff}$ values with the EAGLES code, we derived a lower age limit of 220 Myr. While the Li\,I method is not well-calibrated in this regime, and we caution that this inference is based only on a tentative detection, the result is consistent with the age derived from gyrochronology. Finally, the enhanced activity of the host star ($\log(R'_\mathrm{HK}) = -4.33\pm0.08$) also appears consistent with the young inferred age for this K dwarf \citep{Mamajek2008, Ye2024}.

According to the NASA Exoplanet Archive \citep{christiansen2025nasaexoplanetarchiveexoplanet}, only seven close-in ($P\lesssim10$~d) Neptunes with measured masses and well-constrained young ages ($\leq$ 1.5 Gyr) are currently known. Many reside in young stellar associations, and are ``temporary Neptunes,'' i.e. they are currently inflated by the heat of formation and will cool, contract, and/or evaporate into sub-Neptunes over time \citep[e.g.][]{David2019, Livingston2026}. Planets like V1298\,Tau\,c ($R_p = 5.1\,R_{\oplus}$; \citealt{Livingston2026}) and AU\,Mic\,b ($R_p = 4.2\, R_{\oplus}$; \citealt{Plavchan2020}) therefore cannot be used as young analogs for the present-day Neptune-sized planet population, despite their current Neptune-like sizes. TOI-4129\,b, on the other hand, is old enough that it is unlikely to evolve into a sub-Neptune, but young enough to still be an interesting laboratory for the evolutionary processes active in the first $\sim$Gyr of a Neptune-sized planet's life. 

\subsection{TOI-4129\,b in Context of the Neptune Ridge}

\begin{figure}[t]
\includegraphics[width=\columnwidth]{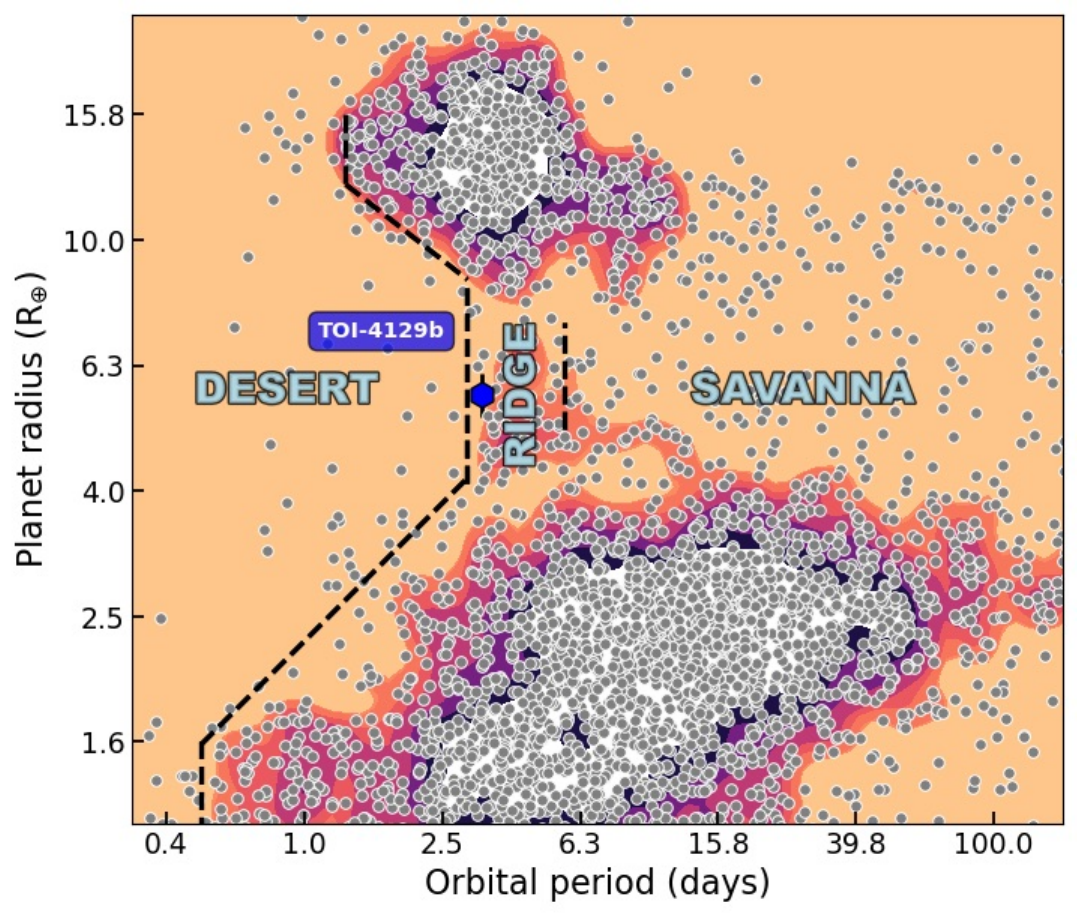}
\caption{The Neptunian landscape with TOI-4129\,b indicated in blue. Boundaries between the desert, ridge, and savanna, as defined by \cite{Castro_Gonz_lez_2024}, are indicated by black dashed lines. TOI-4129\,b is located at the boundary of the Neptune ridge. 
\label{ridge}}
\end{figure}

\begin{figure*}[t]
\centering
\includegraphics[width=\textwidth]{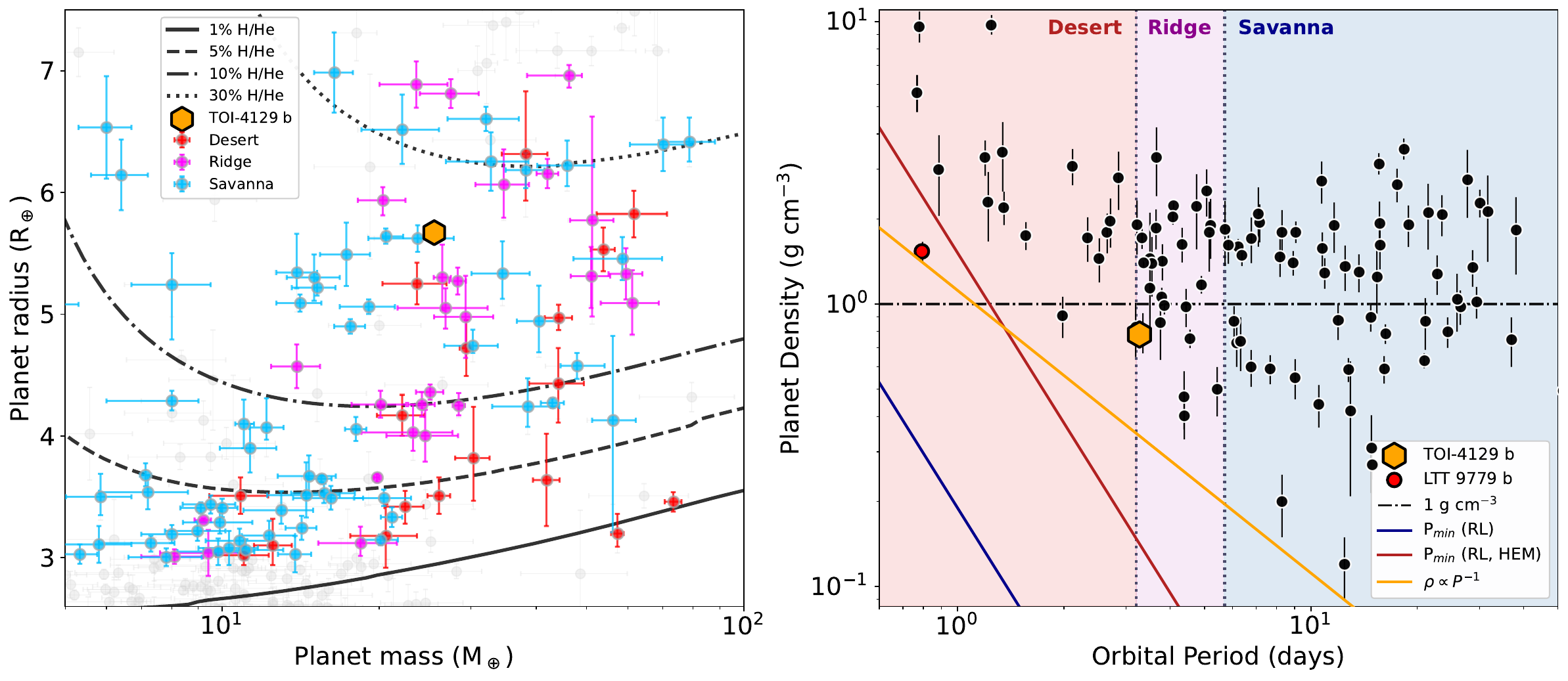}
\caption{TOI-4129\,b represented in the mass-radius and density-period space. Left: The mass-radius distribution of all Neptunian planets with masses and radii with precisions better than 20\%. The isocomposition curves in mass-radius space for several H/He fractions are overplotted using the 1000 K grid from \citet{skinner2026validatedlowtointermediatemassplanetary}. Right: The densities of close-in Neptune planets from the same sample mapped across the Neptunian landscape. The present-day Roche limit \citep{Rappaport_2021} and the HEM Roche limit are denoted $P_\mathrm{min}$ (RL) and $P_\mathrm{min}$ (RL, HEM) respectively. We also note a density-period boundary for Neptunes at $\rho \propto P^{-1}$, first noted by \citet{Bourrier2025}. Both diagrams were generated with \textsf{mr-plotter} \citep{2023A&A...675A..52C}, using data from the NASA Exoplanet Archive \citep{christiansen2025nasaexoplanetarchiveexoplanet}.
\label{MR+PD}}
\end{figure*}

With a period of $3.27$~d, TOI-4129\,b joins the population of planets in the Neptune ridge, shown in Figure~\ref{ridge}. We found that TOI-4129\,b has a radius of $5.7\pm0.2\,R_\oplus$ and a mass of $25.5\pm6.3\,M_\oplus$, giving it a fairly low density of $0.8\pm0.2$\,g\,cm$^{-3}$. Correspondingly, we find that TOI-4129\,b is gas-rich. Using the mass-radius relations from \citet{skinner2026validatedlowtointermediatemassplanetary}, we find that TOI-4129\,b has an H/He fraction between 10\% and 30\%, which we show on the left-hand panel of Figure~\ref{MR+PD}. On the right-hand panel of this figure, we show TOI-4129\,b in the context of all systems with well-measured densities as a function of orbital period. 

TOI-4129\,b is among the lowest-density Neptunes within $P<4$~d. TOI-4129\,b's relatively low density (compared to other ridge residents, which more typically have $\rho>1$~g~cm$^{-3}$) may be a consequence of its youth. At 1.2~Gyr in age, the planet could plausibly still be thermally contracting or losing significant mass to atmospheric photoevaporation \citep{2017ApJ...847...29O}. Additional heating from tides could also contribute to maintaining an inflated radius \citep{Leconte2010, Sethi2025}. Our fits permit orbital eccentricities of $e < 0.2$ (95\% confidence), and young planets like TOI-4129\,b may exhibit non-zero eccentricities because tidal circularization may not yet be completed.

Figure~\ref{MR+PD} also demonstrates an interesting property of the population. From desert to ridge to savanna, Neptunes appear to respect a universal lower boundary in period-density space. This ``density brink'' was first noted by \citet{Bourrier2025}. The Roche limit also gives rise to a period-density boundary \citep[e.g.][]{Rappaport_2021}, but it is inconsistent with the observed boundary. The Roche limit boundary moves outward when assuming that these planets migrated via HEM \citep[and thus we are observing them at $a_\mathrm{final}\approx2r_\mathrm{peri}$;][]{Ford2006} and applying the constraint that the planets should not be tidally disrupted during migration, but the HEM Roche limit still does not quite match up with the period-density boundary of the population. Interestingly, we find that LTT 9779\,b \citep{Jenkins2020} sits well below the HEM curve. If this planet underwent HEM, it must have subsequently undergone significant tidal orbital decay \citep[e.g.][]{Hallatt2026}; otherwise, if its present position reflects $a_\mathrm{final}$, it would have been disrupted during HEM. A recent analysis by \citet{Castro-Gonzalez2026} similarly finds a few Neptunes below the HEM Roche limit curve, although their analysis adopts a higher $\eta = 4$ (treating it as a fitting parameter for the Neptune ridge population). In general, the hot Neptune $\rho-P$ boundary is not quite as steep as predicted by the Roche limit, and another physical mechanism may be responsible. Future discoveries of (especially low-density) Neptune-sized exoplanets may help reveal the nature of the density brink. 

\subsection{Evolutionary Simulations of TOI-4129\,b}

We simulated the thermal evolution and atmospheric mass loss of TOI-4129\,b under the evolving stellar irradiation with the \textsf{JADE} code \citep{Attia2021,Attia2025}\footnote{\url{https://gitlab.unige.ch/spice_dune/jade}}, following the procedure detailed in \cite{Bourrier2025}. First, a rotating stellar model representative of TOI-4129 was computed with the Geneva stellar evolution code (GENEC; \citealt{Eggenberger2008}) using stellar radius and effective temperature constraints given in Tables~\ref{tab:stellar} and \ref{tab:orbital}. This model accounts for the internal transport of angular momentum by hydrodynamic and magnetic instabilities (see e.g., \citealt{Eggenberger2022}). Based on the structural and rotational profile from this model, the evolution of the high-energy fluxes emitted by the host star is then computed (see \citealt{Pezzotti2021}). Due to the relatively fast surface rotation rate observed for this star, we decided to adopt the case of fast rotators with an initial rotation rate of $18\Omega_{\odot}$ and a duration of the disk-locking phase of 2~Myr (see \citealt{Eggenberger2019}). This rotational history then results in a given evolution of the X-ray luminosity of the host star and to a corresponding extreme ultraviolet (EUV) luminosity computed with the relation between the X-ray and EUV luminosity provided by \cite{Sanz-Forcada2011}. 

We then simulated the internal structure of TOI-4129\,b with \textsf{JADE}. The planetary structure consists of an iron core ($\alpha$-Fe), a silicate mantle (MgSiO$_3$), and an H/He-dominated envelope. The latter is divided into an upper region absorbing the stellar radiation and a lower region redistributing energy via radiation and convection. The Rosseland mean opacity of the envelope (computed from the tabulated opacities of \citealt{Ferguson2005}) can be increased by including trace amounts of metals, controlled by their mass fraction relative to the envelope. \textsf{JADE} fits for the mantle-to-planet mass fraction, envelope-to-planet mass fraction, and metal-to-envelope mass fraction $Z_{\rm env}$ with an MCMC approach, using the measured planet radius as a constraint and the measured planet mass as a prior. For a given set of parameters, the code integrates from the top the 1D thermodynamical structure of the envelope and the polytropic equation of state of the mantle and core, iterating over a grid of radii until finding the one that yields zero mass at planet center. The retrieval is performed at 1.2~Gyr to account for the impact of the present-day stellar bolometric luminosity and planetary internal luminosity on the atmospheric structure. The planet orbital properties were set to their present-day values (Table~\ref{tab:orbital}) and the atmospheric He abundance to solar (27.5\%). Uniform priors ($\mathcal{U}$(0,1)) were set on the atmosphere and mantle mass fractions and a narrower prior ($\mathcal{U}$(0,0.1)) on the metal-to-envelope mass fraction. Atmospheric metallicity and the partitioning of the core and mantle mass are poorly constrained, as expected from the degeneracies inherent to internal structure retrievals, while we can constrain the envelope mass fraction to 19.7$^{+3.1}_{-3.4}$\%. This value is in agreement with the H/He fraction suggested by the \citet{skinner2026validatedlowtointermediatemassplanetary} grids in Figure~\ref{MR+PD}.

\begin{figure}
    \centering
    \includegraphics[width=\linewidth]{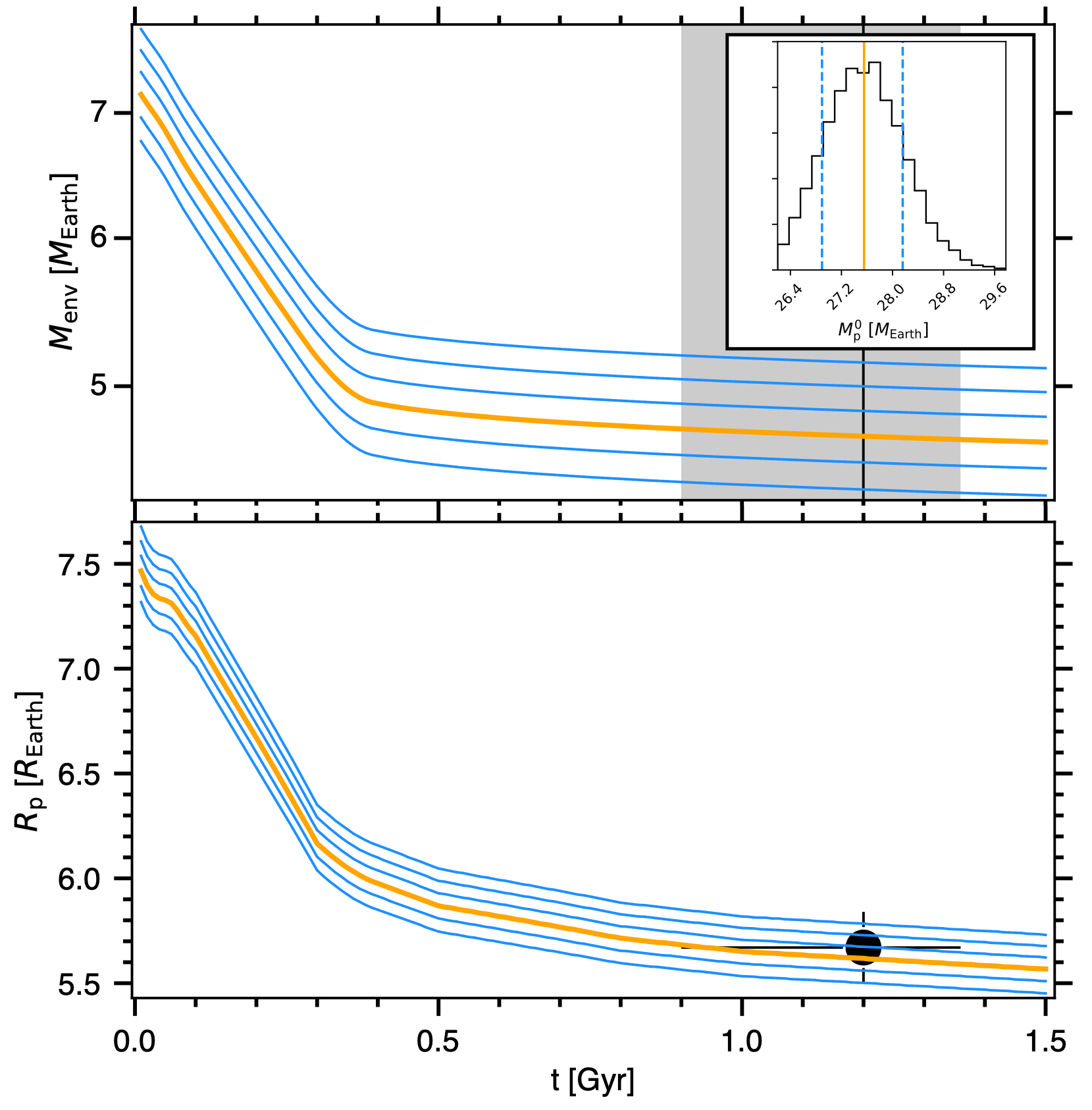}
    \caption{\textsf{JADE} exploration of TOI-4129\,b atmospheric evolution. \textit{Top}: Temporal evolution of the planet envelope mass, in the best-fit simulation (orange curve) and a set of simulations from within the 1$\sigma$ credible interval (blue curves). A vertical black line with shaded band marks the assumed age and uncertainties. The inset shows the PDF of the initial planet mass, with the median highlighted as an orange line and 1$\sigma$ credible interval as dashed blue lines. \textit{Bottom}: Same as the top panel for the planet radius. The black point indicates the measured value.}
    \label{fig:TOI4129_JADE}
\end{figure}

Finally, we ran \textsf{JADE} evolutionary simulations over a grid of initial planet envelope mass, from the expected dissipation of the protoplanetary disk (10~Myr) up to 1.5~Gyr, constraining the retrieval with the present-day planet radius and system age. Other properties were fixed to their values from Tables~\ref{tab:stellar} and \ref{tab:orbital} and to the results of the internal structure retrieval. The stellar luminosity curve was set to the \textsf{GENEC} model, controlling the temperature profile of the planetary envelope together with internal heating, and its erosion under XUV-driven photoevaporation. Results of the simulations are presented in Figure~\ref{fig:TOI4129_JADE}. TOI-4129\,b may have formed with a total initial mass of 27.6$^{+0.6}_{-0.7}$\,$M_\oplus$, corresponding to an envelope mass fraction of about 26\%. About a third of this envelope would then have been lost within the first 300 Myr, when the star was most energetic, after which TOI-4129\,b continued eroding and contracting at a slower rate.

\subsection{Prospects for Future Atmospheric Spectroscopy} \label{subsec:future:atmosphere}

The Transmission Spectroscopy Metric \citep[TSM;][]{2018PASP..130k4401K} quantifies a planet's suitability for transmission spectroscopy, and is used to identify top atmospheric characterization targets for future observation. We found that TOI-4129\,b has a TSM of $112 \pm 30$, which makes it a good candidate for observations with the James Webb Space Telescope (JWST) and the future Atmospheric Remote-Sensing Infrared Exoplanet Large-survey (Ariel), for which it will likely be a Tier 2 target \citep{Edwards_2022}. We therefore proceeded to make a detailed estimate of the potential for atmospheric spectroscopy with Ariel. The Ariel noise model was generated with \textsf{ArielRad} v2.4.26, using \textsf{ExoRad} v2.1.111 as the underlying generic point-source radiometric simulator and the Ariel payload configuration \textsf{Payload} v0.0.17. \textsf{ArielRad} is the Ariel radiometric model developed to evaluate the payload science performance and construct the wavelength-dependent noise budget, while \textsf{ExoRad} provides the instrument-independent radiometric framework used to estimate the signal and noise terms for photometric channels and spectral bins \citep{Mugnai2020ArielRad,Mugnai2023ExoRad}.

We generated and retrieved three cloud-free transmission spectra with \textsf{TauREx} \citep{AlRefaie2021TauREx,AlRefaie2022TauRExChem}, using the \textsf{FastChem} equilibrium-chemistry plugin \citep{Stock2018FastChem}. The forward models used the stellar and planetary properties in Tables~\ref{tab:stellar} and \ref{tab:orbital}. The atmosphere was assumed to be isothermal and cloud-free, with absorption, Rayleigh scattering, and H$_2$--H$_2$/H$_2$--He collision-induced absorption included in the transmission calculation. The chemical calculation included H, He, C, N, O, and S, with C/O fixed to 0.55. We considered injected metallicities of $0.1$, $1$, and $100\,Z_\odot$. In each retrieval we fitted $R_p$, the isothermal temperature, and the atmospheric metallicity using \textsf{MultiNest} with 1000 live points \citep{Feroz2009MultiNest}. The retrievals used a log-uniform atmospheric-metallicity prior extending over $10^{-3}<Z_{\rm atm}/Z_\odot<10^{3}$, so that sub-solar, near-solar, and highly enriched atmospheres were all included in the explored parameter space.

\begin{figure}[t]
	\centering
	\includegraphics[width=\columnwidth]{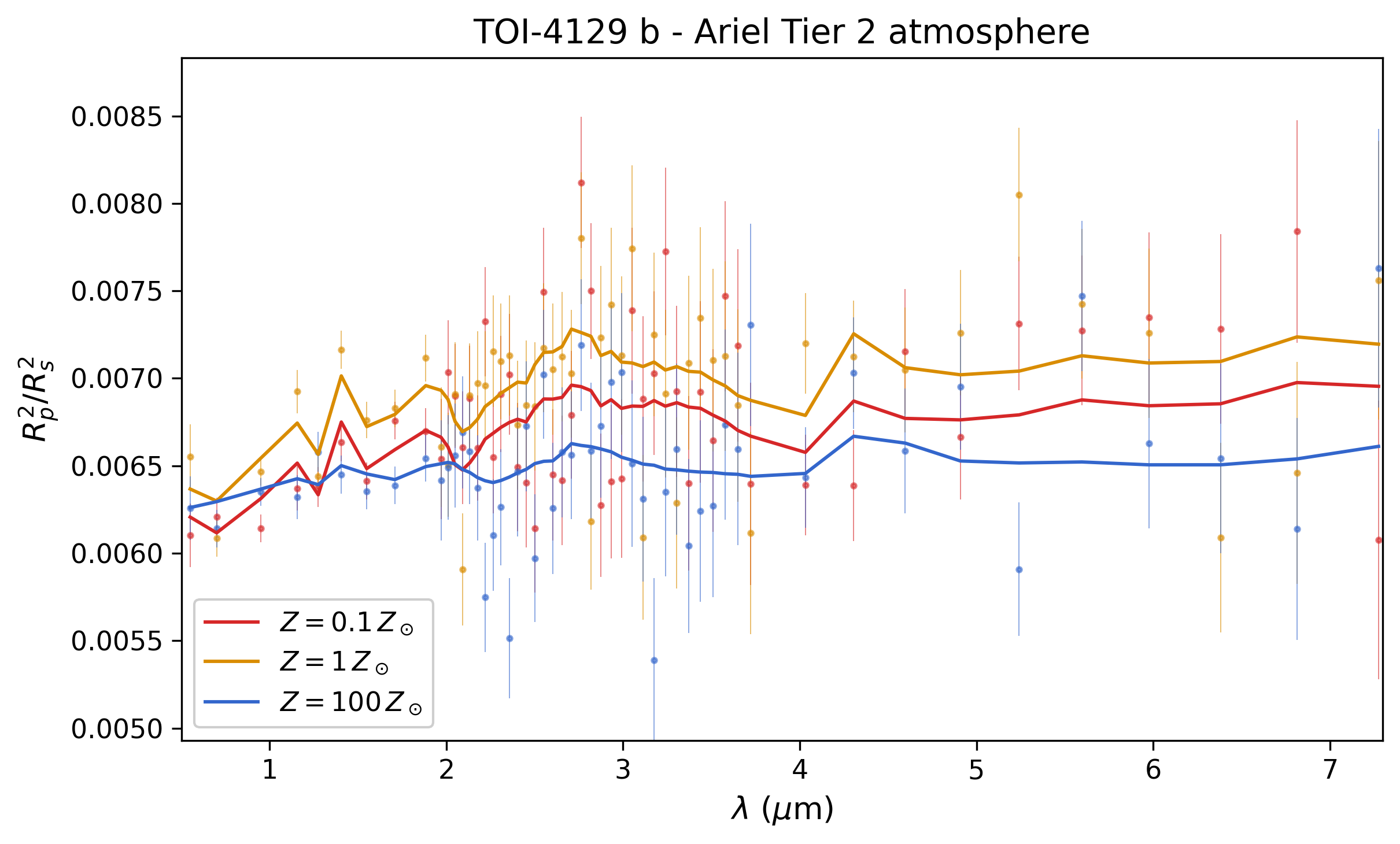}
	\caption{Simulated Ariel Tier~2 transmission spectra for the three \textsf{FastChem} metallicity scenarios considered for TOI-4129\,b. Points with 1$\sigma$ error bars show the binned simulated Ariel observations, while solid curves show the best-fit \textsf{TauREx} spectra. The three cases correspond to injected metallicities of $0.1$, $1$, and $100\,Z_\odot$.}
	\label{fig:ariel_metallicity_spectra}
\end{figure}

Figure~\ref{fig:ariel_metallicity_spectra} shows that Ariel would sample broad molecular structure over the optical and near-infrared, with the largest leverage coming from the $\sim1$--$5~\mu$m region. The three scenarios produce distinguishable spectral morphologies in the adopted cloud-free equilibrium-chemistry models. The highly enriched atmosphere has the most distinct transmission spectrum, as expected from the combined effects of increased molecular opacity and higher atmospheric mean molecular weight, whereas the low- and near-solar-metallicity cases are more similar and require the full wavelength-dependent spectral shape for discrimination.

\begin{figure}[t]
	\centering
	\includegraphics[width=\columnwidth]{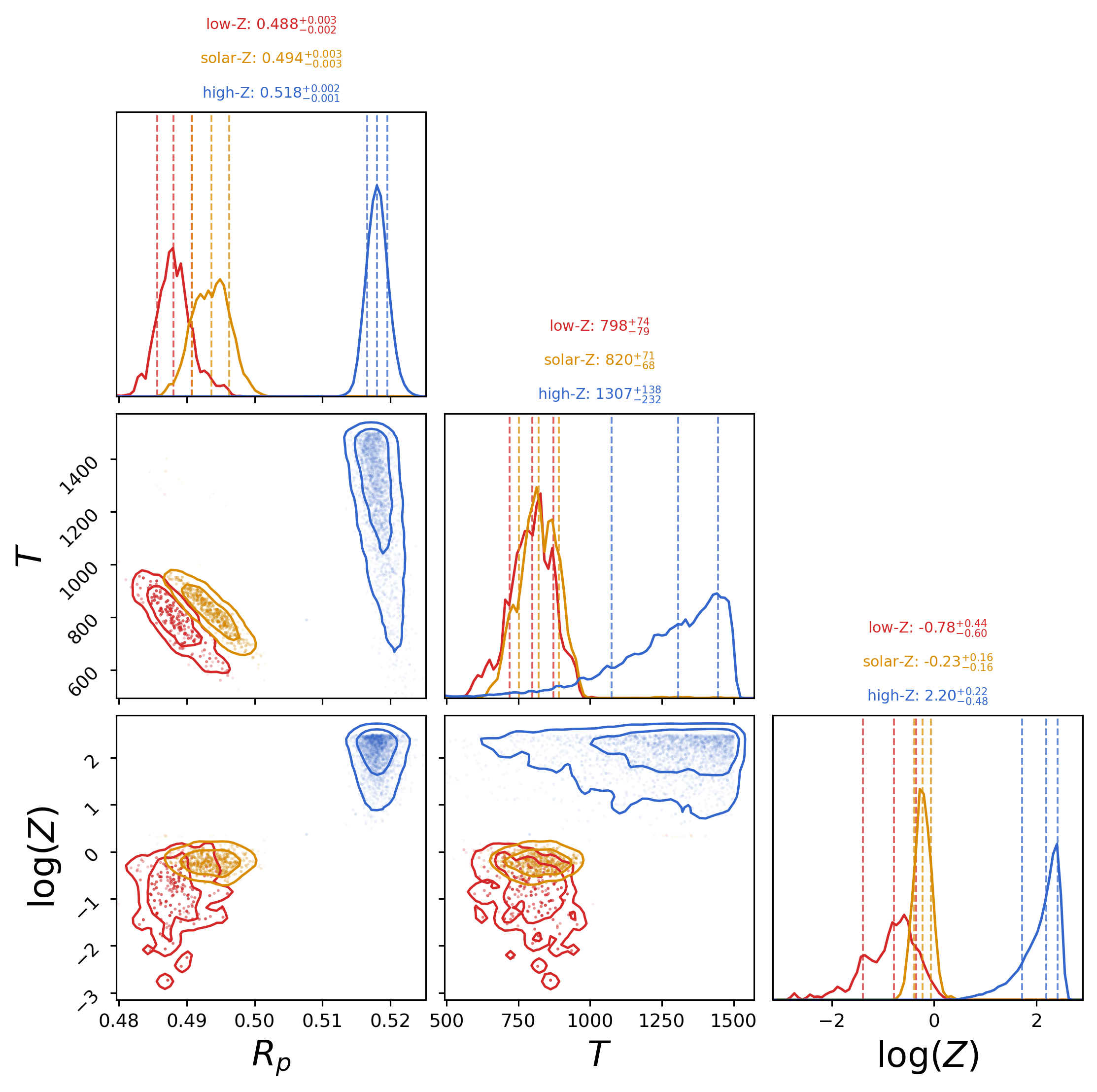}
	\caption{Posterior comparison for the three Ariel/\textsf{TauREx} retrievals. The diagonal panels show the marginalized posterior distributions, with the median and 16th--84th percentile intervals reported for each metallicity scenario. The lower panels show the corresponding two-dimensional posterior contours. Only the retrieved parameters are shown: planetary radius, isothermal atmospheric temperature, and atmospheric metallicity.}
	\label{fig:ariel_metallicity_corner}
\end{figure}

The posteriors in Figure~\ref{fig:ariel_metallicity_corner} demonstrate that Ariel should be able to distinguish the metal-enriched scenario from the solar and metal-poor scenarios. For the injected $0.1\,Z_\odot$ case, the retrieved atmospheric metallicity is $Z_{\rm atm}=0.165^{+0.286}_{-0.124}\,Z_\odot$, corresponding to $\log_{10}(Z_{\rm atm}/Z_\odot)=-0.78^{+0.44}_{-0.60}$. For the injected solar-metallicity case, the retrieval gives $Z_{\rm atm}=0.593^{+0.271}_{-0.180}\,Z_\odot$, or $\log_{10}(Z_{\rm atm}/Z_\odot)=-0.23^{+0.16}_{-0.16}$. For the injected high-metallicity case, the retrieval gives $Z_{\rm atm}=158^{+103}_{-105}\,Z_\odot$, or $\log_{10}(Z_{\rm atm}/Z_\odot)=2.20^{+0.22}_{-0.48}$. Thus, the high-metallicity atmosphere is clearly separated from the low- and near-solar-metallicity solutions, while the low- and solar-metallicity posteriors are partially overlapping but still shifted relative to one another.

This feasibility test indicates that Ariel Tier~2 observations should be able to distinguish a strongly metal-enriched atmosphere for TOI-4129\,b from low- or near-solar-metallicity alternatives under the adopted model assumptions (most notably, that of a cloud-free atmosphere). The current setup fits a bulk atmospheric metallicity rather than individual molecular mixing ratios, assumes chemical equilibrium and an isothermal pressure--temperature profile, and omits clouds and hazes. Within those assumptions, the simulated Ariel spectra carry enough information to test whether this young Neptune-ridge planet has a substantially metal-enriched envelope, strengthening the case for TOI-4129\,b as a good target for future atmospheric characterization.

\subsection{Prospects for Future Obliquity Constraints} \label{subsec:future:obliquity}

Measurements of stellar obliquity, the angle between a star’s rotation axis and the orbital axis of its planets, can provide valuable insight into planetary formation and migration histories. While planets that migrate smoothly through a protoplanetary disk are generally expected to remain aligned with their host stars \citep[though see][]{Biddle2025}, dynamical processes such as planet-planet scattering or HEM can produce substantial spin-orbit misalignment. Recent studies of Neptune-sized planets have revealed a diverse obliquity distribution of aligned, misaligned, and retrograde systems, suggesting that multiple migration pathways contribute to the observed population \citep[e.g.][]{Bourrier2018,EspinozaRetamal2026, Mantovan2026}. However, because TOI-4129\,b is a young planet, it has not had much time to tidally realign, and realignment timescales for Neptunes are longer than the age of the Universe anyway \citep[e.g.][]{Mantovan2024, EspinozaRetamal2026, Hallatt2026}. Thus, by measuring the obliquity of this planet, we can directly access a Neptune's primordial spin-orbit (mis)alignment. Moreover, our rotation period will eventually allow for a constraint on TOI-4129\,b's three-dimensional obliquity angle.

To begin, we made a rough estimate of the classical Rossiter-McLaughlin (RM) amplitude \citep{1924ApJ....60...15R, McLaughlin1924} using the equation presented in 
\citet{Gaudi_2007}, finding it to be approximately $7.4 ~\textrm{m}~\textrm{s}^{-1}$ (assuming a $v\sin i=2$~km~s$^{-1}$, consistent with the $v_\mathrm{eq}$ inferred from our rotation period and $R_\star$ and the direct $v\sin i$ upper limit reported in Table~\ref{tab:stellar}). This precision has been achieved for RM measurements before, but the host star is quite faint ($V=12.9$), so we proceeded to make a more realistic calculation of the signal detectability. 

We simulated the RM effect observability for the Keck Planet Finder \citep[KPF;][]{Gibson2024}. First, we generated mock transit datasets of the system with the \textsf{ANTARESS} pipeline \citep{Bourrier2024} using the properties in Tables~\ref{tab:stellar} and \ref{tab:orbital}, separately considering aligned and polar configurations. We considered exposure times of 213 and 530~s (both including 47~s overheads), required to reach a minimum and optimal S/N of 30 and 50 on the stellar spectrum (typical values to ensure proper reduction of transit spectra). For both exposure times we scaled the flux level of the mock disk-integrated cross-correlation functions (CCFs) in each assumed orbital configuration to match the precisions expected with KPF on the derived RVs (4.3 and 1.7 m~s$^{-1}$, respectively, determined using the KPF Exposure Time Calculator\footnote{https://github.com/california-Planet-Search/KPF-ETC}). The resulting simulated timeseries contained about 27 and 11 in-transit exposures, respectively, plus enough out-of-transit exposures to build a high S/N stellar line master. Then, we processed these mock datasets with \textsf{ANTARESS} similarly to how we would process real observations. We extracted planet-occulted CCFs which we then analyzed using the RM Revolutions technique \citep{Bourrier2021}.

\begin{figure}[t]
\includegraphics[width=\columnwidth]{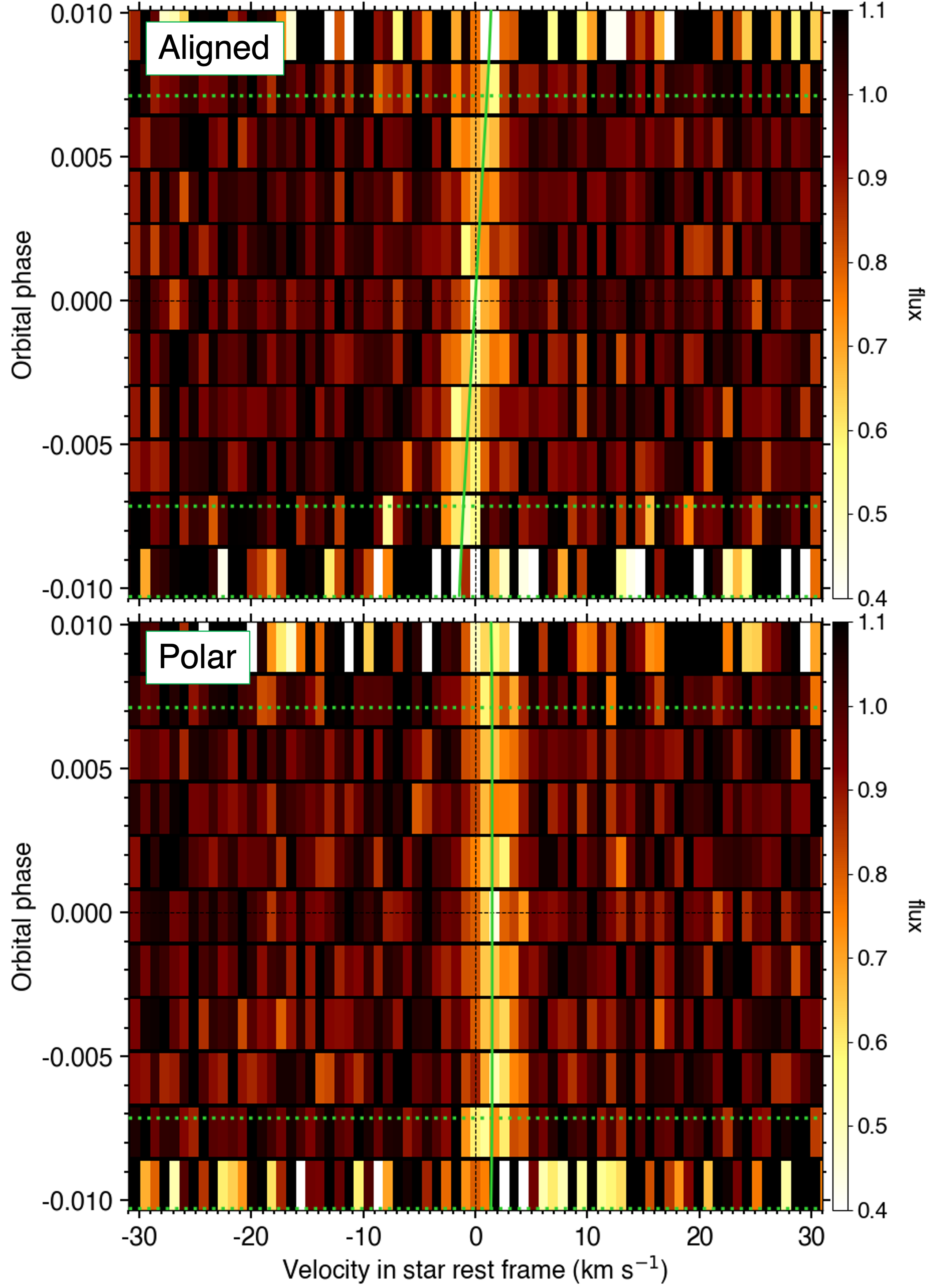}
\caption{Mock datasets of planet-occulted CCFs simulated for KPF with an exposure time of 530\,s in an aligned or polar configuration. The local stellar line profile is clearly detectable as a bright streak, with sufficient S/N to significantly constrain the sky-projected angle and stellar rotational velocity for any orbital architecture.
\label{RMR_predic}}
\end{figure}

The sky-projected spin-orbit angle and rotational velocity are retrieved within $\sim$1$\sigma$ in all four cases. For both exposure times, we expect a better precision on $\lambda$ in the aligned compared to the polar configuration, and conversely for $v\sin{i}$. Compared to 213\,s, an exposure time of 530\,s improves uncertainties on $\lambda$ from $\sim$6$^{\circ}$ (resp. 12$^{\circ}$) to $\sim$4$^{\circ}$ (resp. 8$^{\circ}$) in the aligned (resp. polar) case by providing a higher observational duty cycle while adequately sampling the transit (Figure~\ref{RMR_predic}). A shorter exposure time may however be useful to search for finer stellar surface effects such as differential rotation or convective blueshift \citep[e.g.][]{Cegla2016}.

In all, we find that TOI-4129\,b is a suitable target for RM observations. Given the faintness of the host star ($V=12.9$), we expect that stable spectrographs on the largest telescopes are necessary to measure the obliquity of TOI-4129\,b. Although these resources are often oversubscribed, determining the primordial three-dimensional obliquity for a relatively young planet in the Neptune ridge would be very scientifically valuable.

\section{Conclusion} \label{sec:conclusion}

We confirmed the discovery of TOI-4129\,b, a member of the ``Neptune ridge'' \citep{Castro_Gonz_lez_2024} orbiting a K dwarf every 3.27~d. We found that the planet has a relatively low density, with a radius of 5.67 $\pm~ 0.17~R_\oplus$, a mass of 25.5 $\pm~6.3~M_\oplus$, and thus a density of 0.78 $\pm$ $0.20\,\textrm{g}\,\textrm{cm}^{-3}$. We find a clear 12.7~d rotation period for the host star in the TESS data, and conclude that the system is fairly young, with a gyrochronal age of 1.20$^{+0.16}_{-0.30}$~Gyr. This is one of the first adolescent planets within the Neptune ridge with a mass constraint. While placing this planet within the context of the Neptunian landscape, we noticed that the Neptunes universally appear to respect a lower boundary in density-period space. Further efforts to discover low-density Neptunes will elucidate the existence and origin of this apparent ``density brink'' \citep{Bourrier2025}. 

Planets at $\sim$Gyr ages exist in a brief but critical evolutionary epoch, where tidal circularization and obliquity damping have not yet concluded, possibly allowing for deeper dynamical insights into the formation and evolution of older planets. In this context, we found that TOI-4129\,b is a suitable target for RM observations, and given the well-measured stellar rotation period we expect to be able to determine a 3D obliquity for this system, although this will require a high-precision stabilized spectrograph on a large telescope. We also found that TOI-4129\,b has a TSM of 112 $\pm~30$, making it a good target for future atmospheric observations with JWST and Ariel. Together, future atmospheric and dynamical constraints on this adolescent Neptune offer a valuable opportunity to improve our understanding of how Neptunian planets form and evolve.

\nocite{https://doi.org/10.26133/nea2,https://doi.org/10.26134/exofop3,https://doi.org/10.17909/3y7c-wa45,https://doi.org/10.17909/fwdt-2x66,https://doi.org/10.17909/t9-nmc8-f686, https://doi.org/10.26131/IRSA598}

\begin{acknowledgments}

We thank Nicholas Dowling for obtaining the Wendelstein light curve and uploading it publicly on ExoFOP. S. L. was supported by the Carnegie Astrophysics Summer Student Internship (CASSI) and thanks the entire CASSI cohort.

L. N. acknowledges financial contribution from the INAF Large Grant 2023 ``EXODEMO''. D. W. L. thanks SAO for supporting operations of TRES at FLWO. K. A. C. acknowledges support from the TESS mission via subaward s3449 from MIT. Funding for K. B. was provided by the European Union (ERC AdG SUBSTELLAR, GA 101054354). A. C. C. acknowledges support from UKRI/ERC Synergy Grant EP/Z000181/1 (REVEAL). C. R. and P. E. acknowledge support from the FNS grant no 219745 (Asteroseismology of transport processes for the evolution of stars and planets). M. L-M. is supported by individual research time under NASA contracts NAS5-26555 and NAS5-03127 to the Associated Universities for Research in Astronomy for the operation of the Hubble Space Telescope and James Webb Space Telescope Science Operations Centers at STScI. F. Murgas acknowledges the financial support from the Agencia Estatal de Investigaci\'{o}n del Ministerio de Ciencia, Innovaci\'{o}n y Universidades (MCIU/AEI) through grant PID2023-152906NA-I00. A. M. acknowledges a UK Science and Technology Facilities Council (STFC) small grant ST/Y002334/1. A. A. J. and A. M. acknowledge funding from a UKRI Future Leader Fellowship, grant number MR/X033244/1. M. P. acknowledges support from ASI-INAF agreement no. 2025-10-HH.0 ``Partecipazione Italiana al Gaia DPAC – Supporto alle attività di responsabilità del team scientifico''. The work of I. A. S. was conducted under the state assignment of Lomonosov Moscow State University.

Funding for the TESS mission is provided by NASA's Science Mission Directorate. This work is based on observations made with the Italian Telescopio Nazionale Galileo (TNG) operated by the Fundaci\'{o}n Galileo Galilei (FGG) of the Istituto Nazionale di Astrofisica (INAF) at the Observatorio del Roque de los Muchachos (La Palma, Canary Islands, Spain); programs A48TAC\_34 (PI: Vissapragada) and A49TAC\_43 (PIs: Naponiello and Vissapragada). This article is based on observations made with the MuSCAT2 instrument, developed by ABC, at Telescopio Carlos S\'{a}nchez operated on the island of Tenerife by the IAC in the Spanish Observatorio del Teide. Some of the data presented herein were obtained at Keck Observatory, which is a private 501(c)3 non-profit organization operated as a scientific partnership among the California Institute of Technology, the University of California, and the National Aeronautics and Space Administration. The Observatory was made possible by the generous financial support of the W. M. Keck Foundation. The authors wish to recognize and acknowledge the very significant cultural role and reverence that the summit of Maunakea has always had within the Native Hawaiian community. We are most fortunate to have the opportunity to conduct observations from this mountain. This article is based on observations made in the Two-meter Twin Telescope sited at the Teide Observatory of the Instituto de Astrofísica de Canarias (IAC), that Light Bridges operates in Tenerife, Canary Islands (Spain). The observation time rights (DTO) used for this research were consumed in the PEI `TTVARIAT26'. This article used flash storage and GPU computing resources as Indefeasible Computer Rights (ICRs) being commissioned at the ASTRO POC project that Light Bridges operates in the Island of Tenerife, Canary Islands (Spain). The ICRs were consumed in the PEI `TTVARIAT26' with the collaboration of Bechtle and LENOVO. Based on observations obtained with the Samuel Oschin Telescope 48-inch and the 60-inch Telescope at the Palomar Observatory as part of the Zwicky Transient Facility project. ZTF is supported by the National Science Foundation under Grants No. AST-1440341 and AST-2034437 and a collaboration including current partners Caltech, IPAC, the Oskar Klein Center at Stockholm University, the University of Maryland, University of California, Berkeley, the University of Wisconsin at Milwaukee, University of Warwick, Ruhr University, Cornell University, Northwestern University and Drexel University. Operations are conducted by COO, IPAC, and UW.

This work is partly supported by JSPS KAKENHI Grant Numbers JP24H00017, JP24K00689, JP25K24620, JP26H01402, and JP26K00755, and JSPS Bilateral Program Number JPJSBP120249910. This work is partly financed by the Spanish Ministry of Economics and Competitiveness through grant PGC2018-098153-B-C31. We acknowledge financial support from the Agencia Estatal de Investigaci\'on of the Ministerio de Ciencia e Innovaci\'on MCIN/AEI/10.13039/501100011033 and the ERDF ``A way of making Europe'' through projects PID2021-125627OB-C32 and PID2024-158486OB-C32. This work is supported by the ERC Grant (ERC Advanced Grant SPEAR, GA 101200674) funded by the European Union. Views and opinions expressed are however those of the author(s) only and do not necessarily reflect those of the European Union or the European Research Council Executive Agency (ERCEA). Neither the European Union nor the granting authority can be held responsible for them. This project has received funding from the European Research Council (ERC) under the European Union's Horizon 2020 research and innovation programme (project {\sc Spice Dune}, grant agreement No 947634). This material reflects only the authors' views and the Commission is not liable for any use that may be made of the information contained therein.\\

This research made use of \textsf{nep-des} (available in \url{https://github.com/castro-gzlz/nep-des}) and \textsf{mr-plotter} (available at \url{https://github.com/castro-gzlz/mr-plotter}). NOIRLab IRAF is distributed by the Community Science and Data Center at NSF NOIRLab, which is managed by the Association of Universities for Research in Astronomy (AURA) under a cooperative agreement with the U.S. National Science Foundation. This research has made use of the Exoplanet Follow-up Observation Program (ExoFOP; DOI: 10.26134/ExoFOP5) website, which is operated by the California Institute of Technology, under contract with the National Aeronautics and Space Administration under the Exoplanet Exploration Program. This research has been partially supported by the NASA United States Contributions to the Ariel Preparatory Science Program (USCAPS) under grant number NASA-80NSSC25K0181. The research has been done as part of the NASA Ariel Science Center (NASC) managed by JPL for NASA and operated at Caltech/IPAC in support of NASA's participation in the ESA mission Ariel.

\facility{TESS, ADS, Adams Observatory (0.6 m), Sanchez, WO:0.4m, TTT, FLWO:1.5m, TNG, Keck:II (NIRC2), SAI, Exoplanet Archive}

\software{\textsf{lightkurve} \citep{2018ascl.soft12013L}, \textsf{jaxoplanet} \citep{jaxoplanet}, \textsf{numpyro} \citep{phan2019composable}, \textsf{numpyro\_ext} \citep{numpyro_ext}, \textsf{arviz} \citep{Martin2026}, \textsf{astropy} \citep{2022ApJ...935..167A}, \textsf{corner} \citep{corner}, \textsf{pandas} \citep{mckinney-proc-scipy-2010}, AstroImageJ \citep{Collins:2017}, TAPIR \citep{Jensen:2013}, \textsf{pyaneti} \citep{Barragan2019, Barragan2022}, \textsf{JADE} \citep{Attia2021, Attia2025}, \textsf{gyro-interp} \citep{Bouma_2023}, \textsf{EXOFASTv2} \citep{2017ascl.soft10003E, Eastman2019}, \textsf{MOOG} \citep{osti_4288566}, \textsf{pyMOOGi} \citep{2017AAS...23021607A}, \textsf{TauREx} \citep{AlRefaie2021TauREx, AlRefaie2022TauRExChem}, \textsf{FastChem} \citep{Stock2018FastChem}, \textsf{MultiNest} \citep{Feroz2009MultiNest}, \textsf{ArielRad} \citep{Mugnai2020ArielRad}, \textsf{ExoRad} \citep{Mugnai2023ExoRad}, \textsf{ANTARESS} \citep{Bourrier2024}}
\end{acknowledgments}

\bibliography{sample701}{}
\bibliographystyle{aasjournalv7}

\end{document}